\documentclass[usenatbib,useAMS,fleqn]{mn2e}

\usepackage{epsf}
\usepackage{amssymb}
\usepackage{amsmath}

\newcommand{\diff}{\ensuremath{\mathrm{d}}}

\newcommand{\gtot}{\ensuremath{\bmath{g}_{\mathrm{tot}}}}
\newcommand{\ggrav}{\ensuremath{\bmath{g}}}
\newcommand{\gmag}{\ensuremath{\bmath{g}_{\mathrm{mag}}}}

\newcommand{\gstar}{\ensuremath{g_{\ast}}}

\newcommand{\vrot}{\ensuremath{\bmath{\Omega}}}
\newcommand{\vpos}{\ensuremath{\bmath{r}}}
\newcommand{\vtang}{\ensuremath{\hat{\bmath{e}}_{\mathrm{t}}}}
\newcommand{\vmag}{\ensuremath{\bmath{B}}}

\newcommand{\srot}{\ensuremath{\Omega}}
\newcommand{\srotc}{\ensuremath{\srot_{\rm c}}}

\newcommand{\Bstar}{\ensuremath{B_{\ast}}}
\newcommand{\Beq}{\ensuremath{B_{\rm eq}}}
\newcommand{\Bm}{\ensuremath{B_{\rm m}}}
\newcommand{\Bmp}{\ensuremath{B^{\dagger}_{\rm m}}}

\newcommand{\pot}{\ensuremath{\Phi}}
\newcommand{\potm}{\ensuremath{\Phi_{\rm m}}}
\newcommand{\potmp}{\ensuremath{\Phi^{\dagger}_{\rm m}}}
\newcommand{\dpot}{\ensuremath{\Psi}}
\newcommand{\dpotm}{\ensuremath{\Psi_{\rm m}}}

\newcommand{\rhom}{\ensuremath{\rho_{\rm m}}}
\newcommand{\rhob}{\ensuremath{\rho_{\rm b}}}
\newcommand{\sigmam}{\ensuremath{\sigma_{\rm m}}}
\newcommand{\sigmamdot}{\ensuremath{\dot{\sigma}_{\rm m}}}
\newcommand{\sigmab}{\ensuremath{\sigma_{\rm b}}}
\newcommand{\sigmastar}{\ensuremath{\sigma_{\ast}}}

\newcommand{\rk}{\ensuremath{r_{\textrm{K}}}}
\newcommand{\dr}{\ensuremath{\xi}}
\newcommand{\dri}{\ensuremath{{\xi}_{\rm i}}}
\newcommand{\drm}{\ensuremath{\gamma}}
\newcommand{\drstar}{\ensuremath{\xi_{\ast}}}

\newcommand{\thetam}{\ensuremath{\tilde{\theta}}}
\newcommand{\phim}{\ensuremath{\tilde{\phi}}}
\newcommand{\incl}{\ensuremath{i}}
\newcommand{\phio}{\ensuremath{\phi_{\rm o}}}

\newcommand{\height}{\ensuremath{h_{\rm m}}}
\newcommand{\heightp}{\ensuremath{h^{\dagger}_{\rm m}}}

\newcommand{\sm}{\ensuremath{s_{\rm m}}}
\newcommand{\smp}{\ensuremath{s^{\dagger}_{\rm m}}}

\newcommand{\mdot}{\ensuremath{\dot{m}}}
\newcommand{\Mdot}{\ensuremath{\dot{M}}}
\newcommand{\Mdten}{\ensuremath{\dot{M}_{-10}}}

\newcommand{\mubs}{\ensuremath{\mu_{\rm \ast}}}
\newcommand{\mubm}{\ensuremath{\mu_{\rm m}}}

\newcommand{\dAs}{\ensuremath{\diff A_{\rm \ast}}}
\newcommand{\dAm}{\ensuremath{\diff A_{\rm m}}}

\newcommand{\Mstar}{\ensuremath{M_{\ast}}}
\newcommand{\Rstar}{\ensuremath{R_{\ast}}}
\newcommand{\eratio}{\ensuremath{\varepsilon_{\ast}}}
\newcommand{\estar}{\ensuremath{\eta_{\ast}}}

\newcommand{\wave}{\ensuremath{\lambda}}
\newcommand{\wavez}{\ensuremath{\lambda_{0}}}
\newcommand{\jw}{\ensuremath{j_{\wave}}}
\newcommand{\jz}{\ensuremath{j_{0}}}
\newcommand{\isurfw}{\ensuremath{\mathcal{I}_{\wave}}}
\newcommand{\isurf}{\ensuremath{\mathcal{I}}}
\newcommand{\imp}{\ensuremath{\Delta}}
\newcommand{\zobs}{\ensuremath{z_{\rm o}}}

\newcommand{\velproj}{\ensuremath{v_{\rm p}}}
\newcommand{\vele}{\ensuremath{v_{\rm e}}}
\newcommand{\velesini}{\ensuremath{\vele \sin \incl}}
\newcommand{\velinf}{\ensuremath{v_{\infty}}}
\newcommand{\velesc}{\ensuremath{v_{\rm esc}}}

\newcommand{\tb}{\ensuremath{t_{\rm b}}}
\newcommand{\tff}{\ensuremath{t_{\rm ff}}}

\newcommand{\sOriE}{$\sigma$~Ori~E}
\newcommand{\tOriC}{$\theta^{1}$~Ori~C}

\newcommand{\gauss}{\ensuremath{\textrm{G}}}
\newcommand{\kgauss}{\ensuremath{\textrm{kG}}}

\newcommand{\kelv}{\ensuremath{\textrm{K}}}

\newcommand{\cm}{\ensuremath{\textrm{cm}}}

\newcommand{\cms}{\ensuremath{\textrm{cm}\,\textrm{s}^{-1}}}
\newcommand{\cmss}{\ensuremath{\textrm{cm}\,\textrm{s}^{-2}}}

\newcommand{\gcmcm}{\ensuremath{\textrm{g}\,\textrm{cm}^{-2}}}

\newcommand{\msun}{\ensuremath{\textrm{M}_{\odot}}}
\newcommand{\msunyr}{\ensuremath{\textrm{M}_{\odot}\,\textrm{yr}^{-1}}}

\newcommand{\yr}{\ensuremath{\textrm{yr}}}
\newcommand{\dy}{\ensuremath{\textrm{dy}}}

\newcommand{\eg}{e.g.}
\newcommand{\ie}{i.e.}
\newcommand{\etc}{etc.}
\newcommand{\viz}{viz.}
\newcommand{\cf}{cf.}

\newcommand{\halpha}{H$\alpha$}

\title[Rigidly Rotating Magnetosphere Model]%
	{A Rigidly Rotating Magnetosphere Model for Circumstellar Emission
	  from Magnetic OB Stars}
\author[R. H. D. Townsend \& S. P. Owocki]
       {R. H. D. Townsend$^{1,2}$ %
        \thanks{Email: rhdt@bartol.udel.edu} \& 
        S. P. Owocki$^{1}$ 
	\\
	$^{1}$ Bartol Research Institute,
	University of Delaware,
	Newark, DE 19716, USA
	\\
        $^{2}$ Department of Physics \& Astronomy, 
        University College London, 
        Gower Street, London WC1E 6BT
	}
\date{%
Received: .................................... 
Accepted: ....................................
}

\pagerange{\pageref{firstpage}--\pageref{lastpage}}
\pubyear{2003}

\begin{document}


\maketitle

\label{firstpage}

\begin{abstract}
We present a semi-analytical approach for modeling circumstellar
emission from rotating hot stars with a strong dipole magnetic field
tilted at an arbitrary angle to the rotation axis. By assuming the
rigid-field limit in which material driven (\eg, in a wind outflow)
from the star is forced to remain in strict rigid-body co-rotation, we
are able to solve for the effective centrifugal-plus-gravitational
potential along each field line, and thereby identify the location of
potential minima where material is prone to accumulate. Applying basic
scalings for the surface mass flux of a radiatively driven stellar
wind, we calculate the circumstellar density distribution that obtains
once ejected plasma settles into hydrostatic stratification along
field lines. The resulting accumulation surface resembles a rigidly
rotating, warped disk, tilted such that its average surface normal
lies between the rotation and magnetic axes. Using a simple model of
the plasma emissivity, we calculate time-resolved synthetic line
spectra for the disk. Initial comparisons show an encouraging level of
correspondence with the observed rotational phase variations of
Balmer-line emission profiles from magnetic Bp stars like sigma~Ori~E.
\end{abstract}



\section{Introduction} \label{sec:intro}

High resolution images of the solar corona provide vivid evidence of
how the complex solar magnetic field can structure and confine coronal
plasma \citep[\eg,][]{ZanMas2003}. In other cool, solar-type stars
similar complex magnetic structuring of a hot corona is inferred
indirectly through rotational modulation of the underlying
chromospheric emission network, and by year-to-decade timescale
modulations thought to be analogues of the solar magnetic activity
cycle \citep[\eg,][]{Wil1978,Bal1995}. By contrast, in a subset of
hotter, early-type (O, B, and A) stars, spectropolarimetric
measurements provide quite direct evidence for relatively strong,
stable, large scale magnetic fields, of order $1$--$10\,\kgauss$, and
generally characterized as a dipole with some arbitrary tilt relative
to the rotation axis. Instead of a hot corona, such stars often
exhibit hydrogen Balmer emission associated with relatively cool
material at temperatures of ca. $20,000\,\kelv$, comparable to the
stellar effective temperature. The present paper develops a
\emph{Rigidly Rotating Magnetosphere} (RRM) model for this emission,
based on the notion that material in the star's radiatively driven
stellar wind is channeled and confined into co-rotating, circumstellar
clouds by a strong, rigidly rotating dipole field.

Following the pioneering detection of strong fields in the chemically
peculiar Ap stars \citep{Bab1958}, observations in the mid- and
late-1970s revealed similar magnetic fields in both the late B-type
helium-weak stars \citep[\eg, a Cen --][]{WolMor1974}, and the earlier
(types B0--B2) helium-strong stars \citep[\eg, \sOriE\
--][]{LanBor1978}. More recently, more moderate magnetic fields have
been detected in Be emission-line stars \citep[\eg, $\beta$~Cep
--][]{Hen2000}, slowly-pulsating B stars \citep[\eg, $\zeta$~Cas
--][]{Nei2003}, and O-type stars \citep[\eg, \tOriC\ --][]{Don2002}.

Many of the Bp stars\footnote{By which we refer to both the
helium-weak and the helium-strong, chemically-peculiar B-type stars.}
exhibit both spectroscopic and photometric variability \citep[see,
\eg,][]{PedTho1977,Ped1979}, strongly correlated with changes in
circular polarization arising from their magnetic fields. These
variations have been interpreted in terms of the same `oblique
rotator' conceptual framework that is applied to the Ap stars:
atmospheric stabilization via a magnetic field allows elemental
diffusion to generate surface abundance anomalies, whose axis of
symmetry is parallel to the magnetic axis, and therefore inclined to
the rotation axis \citep{Mic1981}.

Often, the variability seen in Bp stars manifests itself in
circumstellar as well as photospheric diagnostics. Perhaps the
best-studied example is the B2p helium-strong star \sOriE, which shows
\halpha\ shell-like emission varying on the same $1.19\,\dy$ rotation
period as photospheric absorption profiles and photometric indices
\citep[see][and references therein]{GroHun1982}. From studies of this
star, and from investigations of other Bp stars that exhibit similar
emission \citep[\eg,][]{ShoBro1990,Sho1990}, a common observational
picture has emerged of circumstellar plasma confined into tori or
clouds, and forced into co-rotation, by a strong magnetic field
\citep{Sho1993}.

In the case of \sOriE, the material responsible for both the variable
Balmer emission and the eclipse-like behaviour seen in photometric
light curves \citep[\eg,][]{Hes1976} appears to be concentrated at the
intersection between the rotational and magnetic equators
\citep{GroHun1982,Bol1987,ShoBol1994}. An obvious candidate for
imposing such structure is the centrifugal acceleration arising from
magnetically enforced co-rotation; not only can this force lead to the
required breaking of symmetry about the magnetic axis, it can also
furnish the outward lift necessary for confining plasma toward the
tops of magnetic loops.

This overall scenario is somewhat related to the \emph{Magnetically
Confined Wind Shock} (MCWS) model proposed by
\citet{BabMon1997a,BabMon1997b} to explain the X-ray emission from the
A0p star IQ~Aur and the O7pe star \tOriC. However, their model focuses
on the wind collision shocks that can produce hot, X-ray emitting gas
at the top of closed loops, and does not follow the fate of the
radiatively cooled, post-shock material. Magnetohydrodynamical (MHD)
simulations by \citet{udDOwo2002} and \citet{udD2003} indicate that,
without any rotational support, this material simply falls back along
the field line to the loop footpoint. Nonetheless, recent MHD
simulations of the MCWS scenario applied to \tOriC\ have been quite
successful in reproducing its observed X-ray properties
\citep{Gag2004}.

Unfortunately, a similar MHD simulation test of the
\emph{magnetocentrifugal} confinement scenario is much more difficult
to carry out. The strong magnetic fields characteristic of Bp stars,
coupled with the relatively low densities associated with their lower
mass loss rates, imply a very high Alfv\'{e}n speed. As a result, the
time step required to ensure numerical stability, via the
Courant-Friedrichs-Lewy criterion, becomes quite short. This makes it
very expensive to calculate an MHD model spanning the timescales
($\sim$ days) of interest, even for the relatively simple,
two-dimensional axisymmetric case of a dipole aligned with the
rotation axis. For the more general, tilted-dipole case that would
apply to \sOriE\ and other magnetic hot stars, the three-dimensional
nature of the system makes full MHD simulation impractical.

However, in the strong-field limit, an alternative approach becomes
viable. Under the assumption that the field is sufficiently strong so
as to remain completely rigid, the plasma moves along trajectories
that are prescribed \emph{a priori} by field lines that co-rotate with
the star. This reduces the overall three-dimensional modeling of
circumstellar material into a series of one-dimensional problems for
flow evolving under the influence of an effective gravito-centrifugal
potential.

\citet{MicStu1974} used such an approach to model the magnetosphere of
Jupiter, arguing that exospheric material tends to accumulate in
minima of the effective potential, occurring along field lines that
pass near and above the geostationary orbital radius. \citet{Nak1985}
demonstrated how the same approach can be applied to the circumstellar
material of oblique rotator stars such as \sOriE. More recently,
\citet{Pre2004} have presented an alternative formulation of the
strong-field limit, using the condition of force balance tangential to
field lines to map out the complex surfaces on which circumstellar
material can accumulate. Due to the interrelation between force and
potential, the latter treatment is entirely equivalent to the prior
studies based on effective potential minimization.

In the present paper, we use these studies as the foundation on which
we build the RRM model. Sec.~\ref{sec:model} conducts a detailed
review of the effective-potential formulation for the strong-field
limit; this review serves both to establish a more-rigorous footing
for the analyses by \citet{MicStu1974} and \citet{Nak1985}, and as a
basis for the developments presented in subsequent sections. Using
this formulation, we examine how the loci of effective-potential
minima define a likely accumulation surface for circumstellar material
(Sec.~\ref{sec:surfaces}). We then extend our analysis to a full RRM
model for the circumstellar material, including its hydrostatic
stratification around the potential minima (Sec.~\ref{sec:hydstrat}),
its build-up by feeding from the star's wind outflow
(Sec.~\ref{sec:plasma}), and its associated circumstellar line
emission (Sec.~\ref{sec:emission}). The main text concludes
(Sec.~\ref{sec:discussion}) with a comparison of our analyses with
those from previous studies, and with a brief summary of results
(Sec.~\ref{sec:summary}). Finally, the Appendix provides supporting
analyses of the ultimate centrifugal breakout of accumulated material
against the limited confining effect of a finite-strength magnetic
field.

\section{The Strong-Field Limit} \label{sec:model}


\subsection{Basic Principles} \label{ssec:model-basic}

In developing a model for the strong-field limit, we adopt two basic
assumptions. The first is the `frozen flux' condition of ideal MHD, in
which plasma is constrained to move along magnetic field lines. The
second is that these field lines are both rigid and time-invariant in
the frame of reference that rotates at the same angular velocity
\srot\ as the star. Together, these assumptions lead to a picture of
plasma moving along trajectories that are \emph{fixed} in the
co-rotating frame, these trajectories being none other than the
guiding magnetic field lines.

To develop an understanding of how plasma is channeled along the rigid
field lines, let us first consider the case of a solitary parcel
launched ballistically along one such line. The total instantaneous
vector acceleration \gtot\ experienced by this parcel, in the
co-rotating frame, may be broken down into separate components, \viz
\begin{equation} \label{eqn:accel}
\gtot = \ggrav + \gmag - 2 \vrot \times \dot{\vpos} - \vrot
\times (\vrot \times \vpos);
\end{equation}
here, \ggrav\ is the acceleration due to gravity, and \gmag\ is that
due to the magnetic Lorentz force. The last two terms in this
expression arise from the inertial Coriolis and centrifugal forces due
to the rotation of the reference frame; \vrot\ is the vector angular
velocity describing this rotation, with magnitude $|\vrot| = \srot$,
while \vpos\ is the position vector of the plasma parcel, with its
time derivative $\dot{\vpos}$ giving the corresponding velocity
vector.

At any time, the location of the parcel on its respective field line
may be specified by the arc-length distance $s$ from some arbitrary
fiducial point. The temporal evolution of this field line coordinate
is governed by the equation
\begin{equation} \label{eqn:motion}
\frac{\diff^{2} s}{\diff t^{2}} \equiv \ddot{s} = \gtot \cdot \vtang,
\end{equation}
where
\begin{equation} \label{eqn:vtang-v}
\vtang \equiv \frac{\dot{\vpos}}{|\dot{\vpos}|} = \frac{\dot{\vpos}}{\dot{s}}
\end{equation}
is the unit vector tangent to the parcel's trajectory. By our
assumptions above, this trajectory is always directed along the field
line; therefore, this vector is given by
\begin{equation} \label{eqn:vtang-B}
\vtang = \frac{\vmag}{B},
\end{equation}
where \vmag\ is the local magnetic field vector, of magnitude $B
\equiv |\vmag|$.

The equation of motion~(\ref{eqn:motion}) indicates that the dynamics
of the parcel are governed solely by the component of \gtot\ directed
along its trajectory. The components of \gtot\ perpendicular to
\vtang\ have no effect on these dynamics: while they supply the
centripetal acceleration necessary to change the \emph{direction} of
the parcel's space velocity $\dot{\vpos}$, they leave its speed
$\dot{s}$ unchanged. Since both the Coriolis acceleration and the
Lorentz acceleration
\begin{equation}
\gmag \equiv \frac{1}{4\pi} (\nabla \times \vmag) \times \vmag
\end{equation}
are perpendicular to \vtang\ (see eqns.~\ref{eqn:vtang-v}
and~\ref{eqn:vtang-B}), it follows that the equation of motion does
not depend on the appearance of these terms in the
expression~(\ref{eqn:accel}) for \gtot; accordingly, we find that
\begin{equation} \label{eqn:motion-exp}
\ddot{s} = \left[ \ggrav - \vrot \times \left(\vrot \times \vpos
\right) \right] \cdot \vtang.
\end{equation}
The gravitational and centrifugal terms in the brackets arise from
conservative forces; therefore, they may be expressed in terms of the
gradient of an effective potential \pot, such that the equation of
motion becomes
\begin{equation}
\ddot{s} = - (\nabla \pot) \cdot \vtang.
\end{equation}
Recognizing the right-hand-side as the directional derivative of \pot\
along the field line, we have
\begin{equation} \label{eqn:motion-line}
\ddot{s} = -\frac{\diff\pot}{\diff s}.
\end{equation}
This result is very instructive: it tells us that although the plasma
parcel follows a three-dimensional curve $\vpos=\vpos(s)$ through
space, its motion is governed by a potential function $\pot(s)$ that
arises from sampling $\pot(\vpos)$ along this curve. Throughout, we
term this single-variable function the \emph{field line potential}.

If the field line potential $\pot(s)$ exhibits an extremum, so that
\begin{equation} \label{eqn:extremum}
\frac{\diff\pot}{\diff s} \equiv \pot' = 0
\end{equation}
at some point, then the plasma parcel can remain at rest, with no net
forces acting upon it; the components of the gravitational and
centrifugal forces tangential to the field line are equal and opposite,
while those perpendicular to the field line are balanced by the
magnetic tension. Whether the parcel can remain at such an equilibrium
point over significant timescales (\viz, multiple rotation periods)
depends on the nature of the extremum. In the case of a local maximum,
where $\diff^{2}\pot/\diff s^{2} \equiv \pot'' < 0$, the
equilibrium is unstable, and small displacements away from the
extremal point grow in a secular manner. This is what
\citet{udDOwo2002} found in their MHD simulations of wind outflow from
a non-rotating star, in the case of a moderately strong magnetic field
(their $\estar = 10$). Since the effective potential in the absence of
rotation is just the gravitational potential, the tops of magnetic
loops are local maxima of the field line potential. Therefore,
although plasma at loop tops is supported against gravity by magnetic
tension, it is unstable against small perturbations, and -- as the MHD
simulations show -- it eventually slides down one or the other side of
the loop toward the stellar surface.

In the converse situation, where the extremum in the field line
potential is a minimum, with $\pot'' > 0$, the equilibrium is stable:
any small displacement along the local magnetic field line produces a
restoring force directed back toward the equilibrium point. Such
minima represent ideal locations for circumstellar plasma to
accumulate. Because these potential minima can occur on more than a
single field line, the accumulation is not at an isolated point in
space, but rather is spread across one or more surfaces defined by the
loci at which both $\pot' = 0$ and $\pot'' > 0$. Material that
collects on these \emph{accumulation surfaces} forms a magnetosphere
that is at rest in the co-rotating reference frame; when viewed from
an inertial frame of reference, this magnetosphere appears to rotate
\emph{rigidly} with the star, hence the name chosen for the RRM model.

As a brief aside, it is readily demonstrated that the foregoing
potential-based analysis is entirely equivalent to the force-based
formulation presented by \citet{Pre2004}. For instance, the condition
$\pot' = 0$ for an equilibrium point (stable or unstable) may be
expressed in the form
\begin{equation}
\left[ \ggrav - \vrot \times \left(\vrot \times \vpos
\right) \right] \cdot \vmag = \mathbf{0},
\end{equation}
which comes from combining
eqns.~(\ref{eqn:motion-exp}--\ref{eqn:extremum}) with
eqn.~(\ref{eqn:vtang-B}). This expression can be recognized as the
exact same condition of force equilibrium tangential to the local
field line that \citet{Pre2004} impose in their eqn.~(2).

We turn now to examining the effective potential $\pot(\vpos)$, which
determines the potential $\pot(s)$ along each field line. Within the
Roche limit, where the star is assumed to be so centrally condensed
that it may be treated as a point mass, this effective potential is
given by
\begin{equation} \label{eqn:roche-pot}
\pot(\vpos) = -\frac{G\Mstar}{r} - \frac{1}{2} \srot^{2} r^{2}
\sin^{2}\theta;
\end{equation}
here, $G$ is the gravitational constant, \Mstar\ the stellar mass, and
$r$ and $\theta$ are the radial and colatitude coordinates
corresponding to the position vector \vpos, in the spherical-polar
system aligned with the rotation axis. Let us introduce the Kepler
co-rotation radius\footnote{Applied to Earth, the Kepler radius is
equivalent to the orbital radius of a geostationary satellite.}
\begin{equation} \label{eqn:r-kepler}
\rk = \left(\frac{G\Mstar}{\Omega^{2}}\right)^{1/3},
\end{equation}
at which the gravitational and centrifugal forces balance in the
equatorial plane; then \pot\ may be written as
\begin{equation}\label{eqn:potdef}
\pot(\vpos) = \frac{G\Mstar}{\rk} \left( -\frac{1}{\dr} - \frac{1}{2} \dr^2
\sin^{2}\theta \right),
\end{equation}
where $\dr \equiv r/\rk$ is the radial coordinate in units of the
Kepler radius. This latter form is convenient, because the minima of
$\pot(s)$, along each field line, occur in the same location as the
corresponding minima defined by the dimensionless potential
\begin{equation} \label{eqn:dim-pot}
\dpot(\vpos) \equiv \frac{\rk}{G\Mstar}\,\pot(\vpos) = -\frac{1}{\dr} -
\frac{1}{2} \dr^2 \sin^{2}\theta.
\end{equation}
The advantage of working with this dimensionless potential is that it
is \emph{independent} of the rotation rate \srot; as \citet{Pre2004}
demonstrate, a similar conclusion can be reached in the force-based
formulation of the problem. Accordingly, for each magnetic field
configuration, we only need solve once for the accumulation surfaces
where plasma can remain at rest in stable equilibrium; this solution
can then be mapped onto a specific rotation rate by transforming the
radial coordinate from \dr\ back to $r$.

Looking at the form of the dimensionless potential $\dpot(\vpos)$
introduced above, we can identify two regimes. When $r$ is much
smaller than the Kepler radius \rk, such that $\dr \ll 1$, this
potential is spherically symmetric about the origin, and increases
outwards. Conversely, when the distance from the rotation axis greatly
exceeds the Kepler radius, such that $\dr\sin\theta \gg 1$, the
potential is cylindrically symmetric about the same axis, and
decreases outwards. As we demonstrate in the following sections, it is
in this second regime that the field line potential $\pot(s)$, and its
dimensionless equivalent $\dpot(s)$, exhibit the minima near which
circumstellar plasma accumulates.


\subsection{The aligned dipole configuration} \label{ssec:model-align}

\begin{figure}
\epsffile{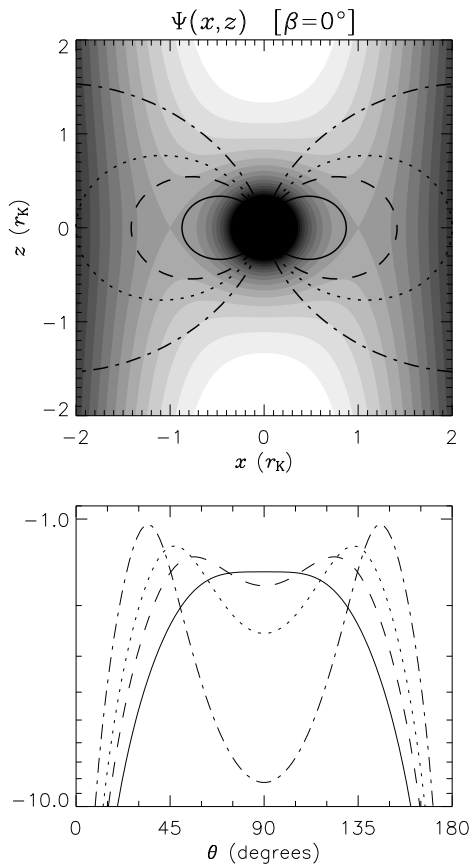}
\caption{A contour map of the dimensionless effective potential over
the $y=0$ plane; white regions corresponds to $\dpot > -0.6$, black
regions to $\dpot < -3$, and the intermediate gray levels are spaced
in increments $\Delta \dpot = 0.15$. Drawn over the map are four
selected field lines for a magnetic dipole aligned with the rotation
axis; these lines have summit radii $\drm=\sqrt[3]{2/3}$ (solid),
$\sqrt{2}$ (dashed), 2 (dotted) and 4 (dot-dashed), and are oriented
at azimuthal angles $\phi=0\degr$ (right) and $\phi=180\degr$
(left). Shown beneath is the dimensionless potential along each of the
$\phi=0\degr$ lines, plotted as a function of $\theta$; note that the
scale of the ordinate is logarithmic.}
\label{fig:roche-align}
\end{figure}

In the foregoing discussion, we argue that circumstellar plasma
accumulates on the surfaces defined by minima of the field line
potential. We now consider the geometry of one such accumulation
surface in the simplest of all configurations, that of a centred
dipole magnetic field aligned with the star's rotation axis. Defining
a Cartesian coordinate system $(x,y,z)$ with origin at the star's
centre and rotation taken along the $z$-axis,
Fig.~\ref{fig:roche-align} shows an $x$ vs. $z$ contour map of the
dimensionless effective potential $\dpot(\vpos)$ (see
eqn.~\ref{eqn:dim-pot}) in the $y=0$ plane. Superimposed over the map
are four curves, each following the parametric equation for a dipole
field line,
\begin{equation} \label{eqn:field-line-align}
\dr = \drm\,\sin^{2}\theta.
\end{equation}
Here, the parameter \drm\ specifies the summit radius (in units of
\rk) of the field line; in the present case of
Fig.~\ref{fig:roche-align}, the $\drm = \sqrt[3]{2/3}$, $\sqrt{2}$, 2
and 4 lines are plotted. The significance of the first value is
discussed below. Beneath the contour map, we plot the dimensionless
potential
\begin{equation}
\dpot(\theta) = -\frac{1}{\drm\sin^{2}\theta} - \frac{1}{2} \drm^{2}
\sin^{6}\theta,
\end{equation}
this being $\dpot(\vpos)$ sampled along the dipole
trajectory~(\ref{eqn:field-line-align}), for each of the four field
lines. For simplicity, we chose the colatitude $\theta$ as the
independent variable in the above expression, rather than the usual
field line coordinate $s$. However, noting that the two are related
via the differential equation
\begin{equation} \label{eqn:arc-len-align}
\frac{\diff s}{\diff\theta} = \rk\, \drm\, \sin\theta \sqrt{1 +
3\cos^{2}\theta},
\end{equation}
it is clear that, everywhere away from the poles, $s$ varies
monotonically with $\theta$.

Inspecting the $\dpot(\theta)$ data for the three outer field lines
($\drm = \sqrt{2},2,4$) shown in Fig.~\ref{fig:roche-align}, minima
can be seen at the stellar equator ($\theta=90\degr$). Since the
aligned dipole configuration is symmetric about the $z$ axis, we can
conclude that the accumulation surface takes the form of an equatorial
disk, with its normal pointing along the rotation axis. However,
because a potential minimum does not occur along the innermost field
line, it is evident that this disk does not extend to the origin, but
instead must terminate at some finite radius. To determine this inner
truncation radius, we observe that
\begin{align} \label{eqn:dpot-2-deriv}
\dpot'' &= \left(
\frac{\diff \theta}{\diff s}\right)^{2} \frac{\diff^{2} \dpot}{\diff \theta^{2}} + 
\frac{\diff \theta}{\diff s} \frac{\diff^{2} \theta}{\diff s^{2}}
\frac{\diff \dpot}{\diff \theta} \\ \nonumber
&= \frac{1}{\rk^{2}} \left(-\frac{2}{\dr^{3}} + 3\right)
\end{align}
for $\theta=90\degr$, where in the second line we make use of the
identity $\dr = \drm$ within the equatorial plane (\cf\
eqn.~\ref{eqn:field-line-align}). We recall that $\dpot''$ must be
positive in order for an extremum ($\dpot'=0$) to constitute part of
an accumulation surface; accordingly, the inner truncation radius is
given by
\begin{equation}
\dri = \sqrt[3]{2/3} \approx 0.87
\end{equation}
at which $\dpot''$ changes from being positive ($\dr > \dri$) to
negative ($\dr < \dri$).

\begin{figure*}
\epsffile{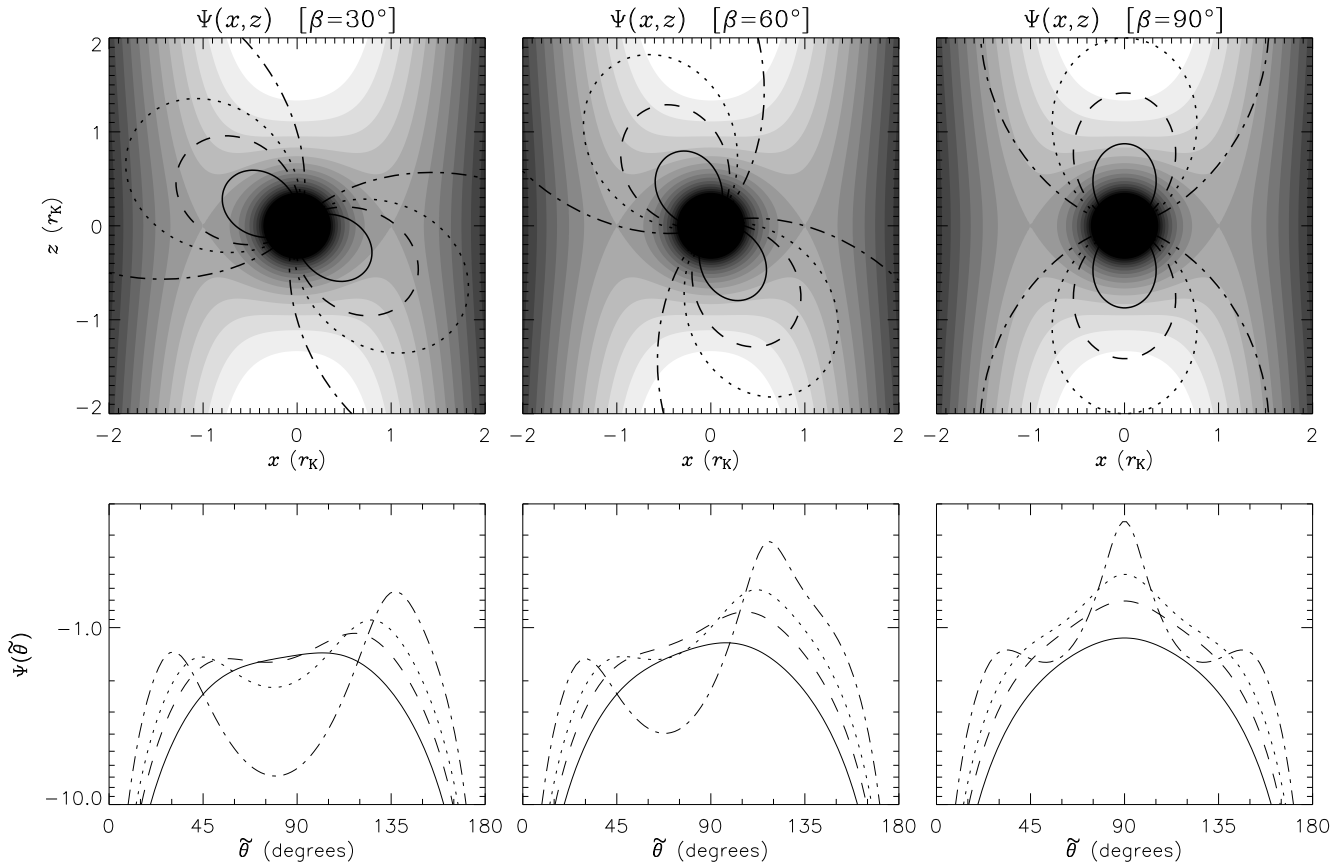}
\caption{As with Fig.~\ref{fig:roche-align}, except that a tilted
dipole field is assumed, at angles $\beta = 30\degr$, $60\degr$ and
$90\degr$ to the rotation axis.} \label{fig:roche-tilt}
\end{figure*}

The reason for our choice of $\drm=\sqrt[3]{2/3}=\dri$ for the
innermost field line in Fig.~\ref{fig:roche-align} should now become
apparent: it ensures that this particular line exactly intersects the
inner edge of the accumulation disk at $\dr = \dri$. Plasma at the
summit of this line is therefore in neutral equilibrium, whereby small
displacements away from the equator produce no net force (to first
order in the displacement) along the field line, either away from the
equilibrium point or toward it. This is evident in the lower panel of
Fig.~\ref{fig:roche-align} from the flatness of this field line's
dimensionless potential at $\theta = 90\degr$.

Throughout the region in the equatorial plane between the truncation
radius $\dr = \dri$ and the Kepler co-rotation
radius\footnote{Apparent in the figure as the twin saddle points at
$(x,z)=(\pm1,0)\,\rk$.} $\dr = 1$, magnetic tension supports
accumulated material against the net inward pull caused by gravity
exceeding the centrifugal force. Beyond this region, when $\xi > 1$,
the centrifugal force surpasses gravity, and the effect of magnetic
tension then becomes to hold material down against the net outward
pull. 

Clearly, the interplay between gravitational and centrifugal forces
has a different significance in the RRM case than it does for a
Keplerian disk; in the latter, material at each radius orbits the star
at a velocity whereby both forces are in exact balance. Such
\emph{complete} force balance is not required in an RRM, inasmuch that
magnetic tension can absorb any net resultant force perpendicular to
field lines \citep{Pre2004}. Only the tangential components of the
forces are required to be in balance, so as to produce an equilibrium
that is stable against small displacements along field lines -- a
point recognized in the original treatment by \citet{MicStu1974},
although these authors employed a more-geometrical approach to arrive
at the same conclusion.


\subsection{The tilted dipole configuration} \label{ssec:model-tilt}

Up until now, we have dealt with the trivial case of a dipole field
aligned with the rotation axis. However, there is nothing that
restricts us to such simple systems; as \citet{Nak1985} first
demonstrated, the rigid-field approach we have presented can be
applied to \emph{arbitrary} magnetic configurations, so long as the
effective potential along each field line can be computed and
minimized.

In the present section we now consider the oblique rotator
configuration, where a dipole is inclined at an angle $\beta$ to the
rotation axis. For such a geometry, eqn.~(\ref{eqn:roche-pot}) still
describes the effective potential in the co-rotating frame, but the
field lines of the tilted dipole now follow the parametric equation
\begin{equation} \label{eqn:field-line-tilt}
\dr = \drm \sin^{2}\thetam,
\end{equation}
where \thetam\ is the colatitude coordinate in the frame of reference
aligned with the magnetic axis. To relate this magnetic reference
frame back to the rotational one, we adopt the convention that the
former is obtained from the latter by rotating by an angle $\beta$
about the Cartesian $y$ axis\footnote{We assume that $\beta > 0$
corresponds to a clockwise rotation, when looking out from the origin
along the positive $y$ axis.}. With this convention, colatitudes
in the two reference frames are related to one another via
\begin{multline}
\sin^{2}\theta = \sin^{2}\thetam\,\sin^{2}\phim + \mbox{}\\
\left(
\sin\beta\,\cos\thetam + \cos\beta\,\sin\thetam\,\cos\phim
\right)^{2},
\end{multline}
were \phim\ denotes the azimuthal coordinate in the magnetic frame.

The latter expression may be used to eliminate the $\sin^{2}\theta$
term from eqn.~(\ref{eqn:dim-pot}), allowing us to express the
dimensionless effective potential along each field line as
\begin{multline} \label{eqn:dim-pot-tilt}
\dpot(\thetam) = -\frac{1}{\drm\sin^{2}\thetam} - \frac{1}{2} \drm^{2}
\sin^{4}\thetam \left[ \sin^{2}\thetam\,\sin^{2}\phim + \mbox{} \right.\\
\left. \left( \sin\beta\,\cos\thetam + \cos\beta\,\sin\thetam\,\cos\phim
\right)^{2} \right];
\end{multline}
here, we have also used eqn.~(\ref{eqn:field-line-tilt}) to eliminate
\dr. It is straightforward to derive an expression for the derivative
$\dpot'$ of this field line potential; however, in contrast to the
aligned dipole configuration, the equation $\dpot' = 0$ for the
extrema of this potential no longer admits algebraic solutions. But in
the $x$--$z$ plane that contains both the magnetic and rotation axes,
we can still illustrate these minima using the same graphical approach
as in Sec.~\ref{ssec:model-align}.

Figure~\ref{fig:roche-tilt} shows similar effective potential plots to
Fig.~\ref{fig:roche-align}, but for dipole field configurations tilted
at angles $\beta = 30\degr$, $60\degr$ and $90\degr$ to the rotation
axis. Focusing initially on the first two cases, we note that a single
potential minimum occurs along the outer three field lines when
$\beta=30\degr$, and along the outer two for $\beta=60\degr$. In each
case, the minima are situated at approximately the same colatitude,
which falls somewhere between the magnetic and rotational equators:
$\thetam \approx 80\degr$ for $\beta = 30\degr$, and $\thetam \approx
70\degr$ for $\beta = 60\degr$. This bisection of the equators arises
because of competition between the two misaligned symmetry axes,
magnetic vs. rotational.

Looking now at the $\beta=90\degr$ case, we can see that \emph{two}
minima -- albeit shallow ones -- occur in the potential along the
outermost field line, at equal distances above ($\thetam \approx
55\degr$) and below ($\thetam \approx 125\degr$) the magnetic equator.
This shows that the accumulation surfaces of tilted configurations can
be significantly more complex than the simple disk found for the
aligned dipole; indeed, as we demonstrate in the following section,
there is no guarantee even that the surfaces are made from a single
contiguous sheet. \citet{Nak1985} overlooked such situations, and it
was not until the study by \citet{Pre2004} that this possibility
became known.

To conclude this section, we draw attention to a limitation of the
graphical approach we have used to illustrate the accumulation
surfaces: although these surfaces are inherently three dimensional,
our approach is restricted to plotting the effective potential and
magnetic field lines over a two dimensional slice through the
system. This is not a problem for the aligned dipole shown in
Fig.~\ref{fig:roche-align}, since rotational symmetry ensures that all
slices containing the polar axis are identical. However, this symmetry
is absent from the tilted field configurations plotted in
Fig.~\ref{fig:roche-tilt}, meaning that the figure cannot indicate the
nature of the accumulation surfaces outside of the $x$--$z$ plane that
contains the misaligned rotation and magnetic axes.


\section{Accumulation Surfaces} \label{sec:surfaces}

\begin{figure*}
\epsffile{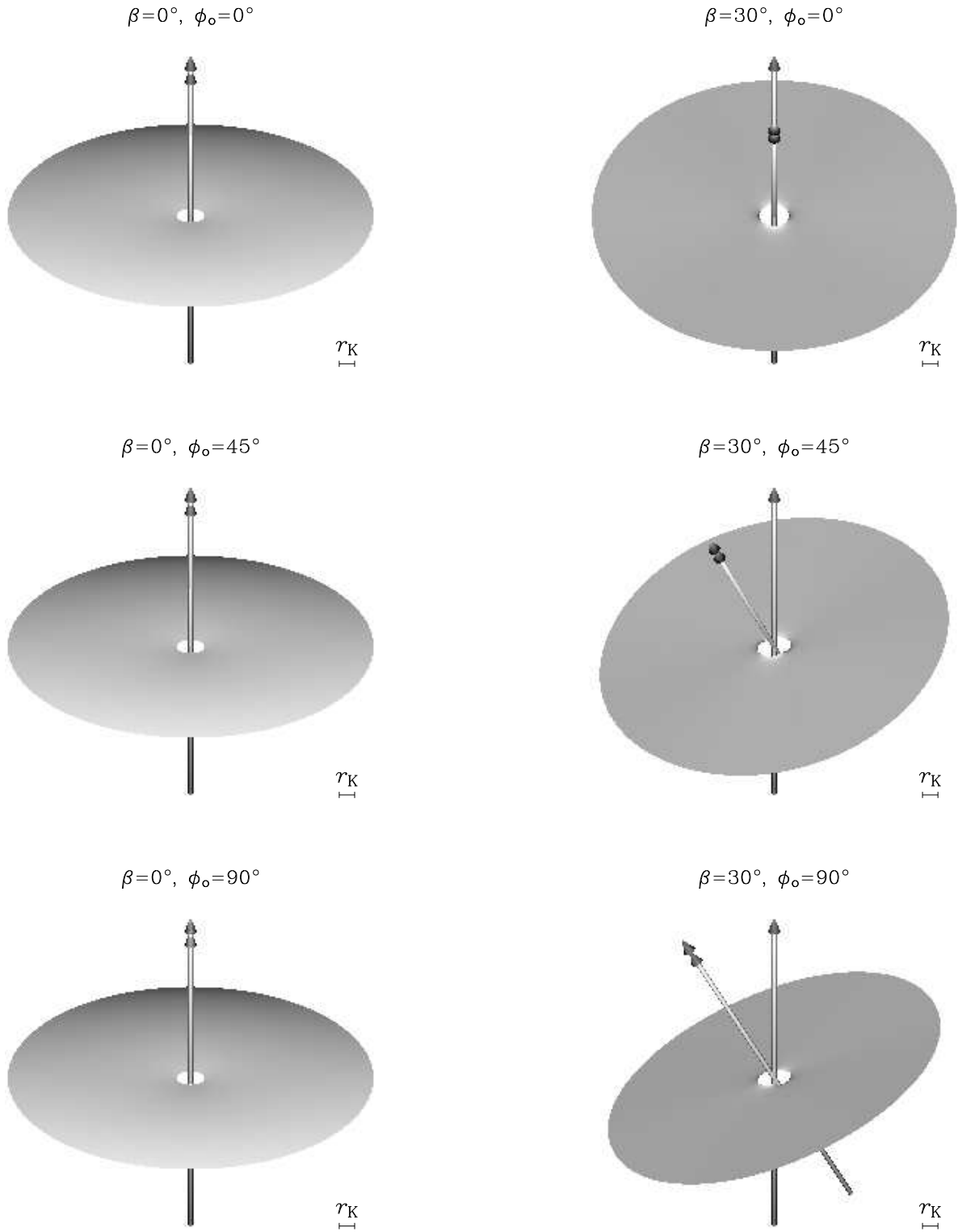}
\caption{\protect Accumulation surfaces for aligned ($\beta=0\degr$;
left-hand column) and tilted ($\beta=30\degr$; right-hand column)
dipole magnetic fields; the surfaces are viewed at an inclination
$\incl=60\degr$ to the rotation axis, and from azimuths
(top-to-bottom) $\phio=0\degr$, $45\degr$ and $90\degr$. In each plot,
the rotation and magnetic axes are shown as single- and double-headed
arrows, respectively; to indicate the scale of the plots, a bar with a
length of one Kepler radius (\rk) is shown in the bottom right-hand
corner of each.} \label{fig:surfaces}
\end{figure*}

\begin{figure*}
\epsffile{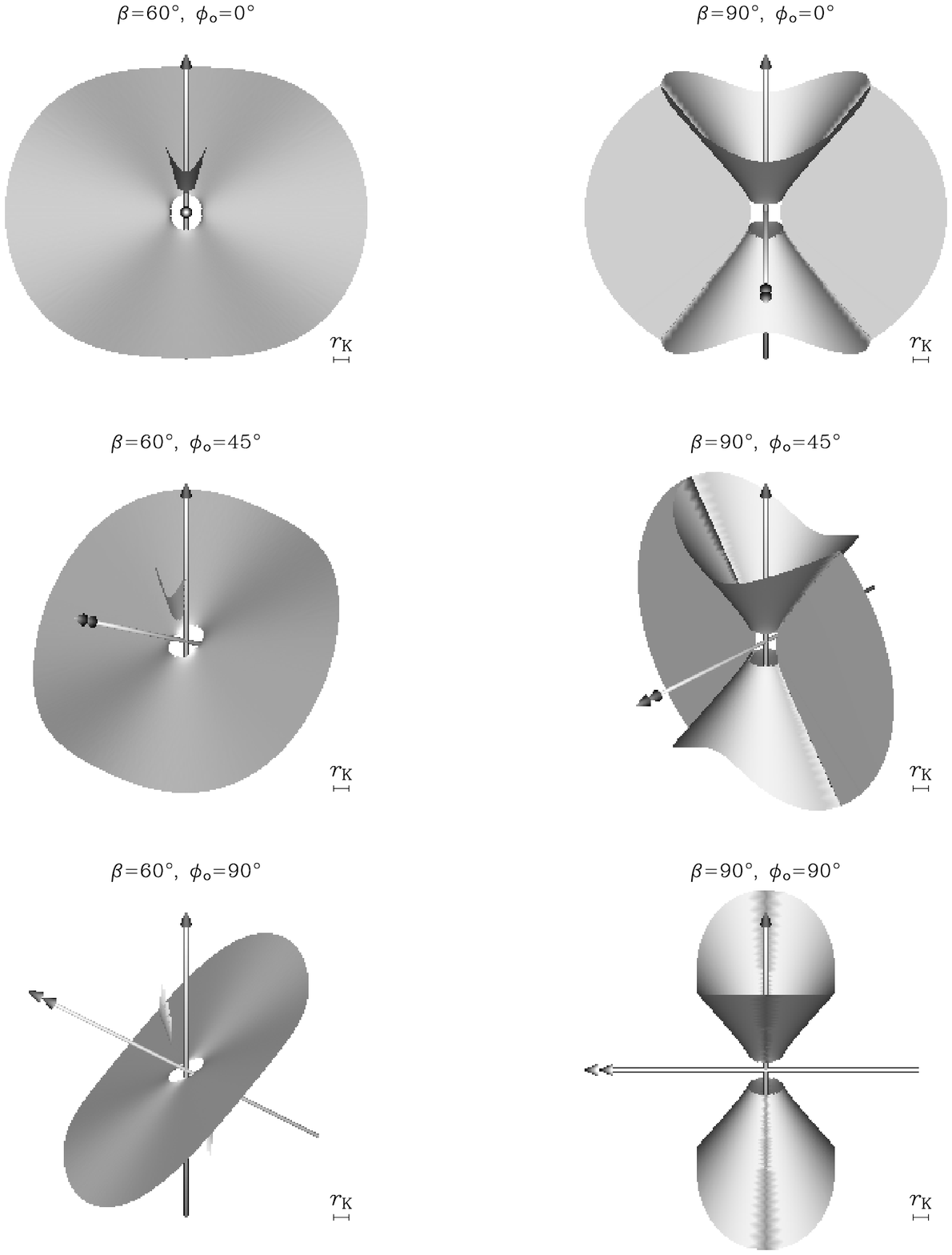}
\contcaption{As before, except that the tilted configurations for
$\beta=60\degr$ (left-hand column) and $\beta=90\degr$ (right-hand
column) are shown.}
\end{figure*}

To overcome these limitations of two-dimensional plots, let us now use
perspective images to show the full three-dimensional form of the
accumulation surfaces. For the same aligned and tilted dipole magnetic
field configurations introduced above, Fig.~\ref{fig:surfaces}
illustrates these as surfaces illuminated by an artificial parallel
light source from the observation point. The vertical, single-headed
arrows denote the rotation axis, while the double-headed arrows shown
in projection at various orientations represent the magnetic axis. For
each of the four values of the tilt angle
$\beta=\{0\degr,30\degr,60\degr,90\degr\}$, the accumulation surface
is shown from three different observation points, situated at the same
inclination $\incl = 60\degr$ to the rotation axis, but having
differing azimuthal angles $\phio = 0\degr$, $45\degr$ and
$90\degr$. Although the full surfaces formally extend to arbitrarily
large radii, for the illustration the outer edge is truncated by
omitting regions threaded by field lines with summit radii greater
than $\drm = 12$.

For the aligned field ($\beta=0\degr$) case, the accumulation surface
is a simple disk lying in the plane of the rotational and magnetic
equators, and so appears identical from all azimuths. The hole at the
centre reflects the lack of potential minima inside the inner
truncation radius \dri\ (Sec.~\ref{ssec:model-align}). For the
$\beta=30\degr$ case the surface is titled, with a mean normal vector
{\it between} the two symmetry axes, in a direction consistent with
the $\thetam \approx 80\degr$ angle found in
Sec.~\ref{ssec:model-tilt}. Although not obvious in
Fig.~\ref{fig:surfaces}, the surface is not strictly planar, but has a
slight warp.

For the greater tilt of the $\beta=60\degr$ configuration, this
warping becomes more apparent. While still shaped approximately like a
disk, the regions nearest the intersection with the plane formed by
the two axes are warped \emph{away} from the rotation axis.
Physically, this arises because the centrifugal force vanishes along
the rotation axis; this force being crucial to forming the potential
minima that make up the accumulation surface, it follows that there is
a `zone of avoidance' around the rotation axis, inside of which the
surface cannot exist. An additional, remarkable feature of the
$\beta=60\degr$ case is the appearance of a pair of secondary
accumulation surfaces, situated in each hemisphere between the
magnetic and rotation axes. These secondary surfaces, which we term
\emph{leaves}, are a consequence of the appearance of an additional
minimum in the effective potential along a particular bundle of
magnetic field lines.

From the analysis in Sec.~\ref{ssec:model-tilt}, it might appear that
such two-minima scenarios are restricted to the perpendicular
($\beta=90\degr$) configuration; however, the appearance of the leaves
in this intermediate-tilt case proves otherwise. In fact, as
\citet{Pre2004} have demonstrated, leaves\footnote{Or `stable chimney
regions' in the parlance of \citet{Pre2004}, the chimney being the
rotation-axis aligned surface defined by $\pot'=0$, that is composed
of both stable \emph{and} unstable equilibrium loci.} occur in
\emph{all} configurations other than the aligned field one, but are
situated at larger and larger radii as $\beta$ decreases toward
zero. In the present case with $\beta=60\degr$, the leaves are
threaded by magnetic field lines for which $\drm \gtrsim 10$; this
explains why the middle panel of Fig.~\ref{fig:roche-tilt}, which
plots field lines up to $\drm=4$, does not exhibit the second
potential minimum associated with the appearance of a leaf.

Turning finally to the perpendicular field ($\beta=90\degr$)
configuration, we see the ultimate product of the disk warping and
leaf formation. The accumulation surface now takes the form of a
partial disk lying in the magnetic equator, intersected by an opposing
pair of truncated cones aligned with the rotation axis. These cones
have an opening half-angle of $\tan^{-1} 2^{-1/2} \approx 35.3\degr$
at asymptotically-large radii, a value that can be derived by setting
the first derivative of $\dpot(\thetam)$ (\cf\
eqn.~\ref{eqn:dim-pot-tilt}) to zero, and then solving for \thetam\ as
$\drm \rightarrow \infty$ \citep[see also][]{Pre2004}. To understand
the unusual geometry of the perpendicular configuration, note that if
$\beta$ were to depart slightly from $90\degr$, then the half of each
cone that lies between the two axes would split off from the main
accumulation surface, and take the form of a leaf. Therefore, we can
recognize the cones as being formed from a merger between the warped
disk and the leaves.

The geometrical complexity of accumulation surfaces, even in the
relatively-simple case of a tilted dipole, was unknown to
\citet{Nak1985}; he had to rely on simple particle-based maps for
visualization (see his figs.~2 \&~3), and was unaware of the
possibility of leaves or of the truncated-cone configuration occurring
at $\beta=90\degr$. Only in recent years have computers become
sufficiently powerful that the visualization of the surfaces is a
relatively straightforward procedure. However, there still remains the
question of how closely a physical system would resemble an
accumulation surface. The answer depends on the nature and
distribution of the matter that populates the surfaces and renders
them visible or detectable. While the accumulation surfaces presented
in this section, and by \citet{Pre2004}, furnish a geometrical picture
of where circumstellar material \emph{can} accumulate, they provide no
indication of how much material \emph{does} accumulate, nor of its
physical conditions -- density, temperature, opacity, emissivity,
\etc. We address these issues in the following sections, where we
derive the distribution of circumstellar gas, and then use this to
calculate the line emission from an RRM model.


\section{Hydrostatic Stratification}\label{sec:hydstrat}

The various processes that could fill effective potential wells with
material may generally be quite dynamic and variable; but over time,
it seems likely that most such material should eventually settle into
a nearly steady, static state. Along any given field line that
intersects with one or more of the potential minima that define
accumulation surfaces, the \emph{relative} distribution of material
density $\rho$ is then set by the requirement of hydrostatic
stratification within the field line potential,
\begin{equation} \label{eqn:hydrostatic}
\frac{\diff p}{\diff s} = -\rho \frac{\diff \pot}{\diff s} \, ,
\end{equation}
where the gas pressure is given by the ideal gas law,
\begin{equation}\label{eqn:gaslaw}
p = \frac{\rho k T}{\mu} \, .
\end{equation}
Taking for simplicity the temperature $T$ and mean molecular weight
$\mu$ to be constant, we can solve for the density distribution along
the field line as
\begin{equation} \label{eqn:density}
\rho(s) = \rho_{m} \,e^{-\mu [\pot(s) - \potm]/k T} \, ,
\end{equation}
where $\rhom \equiv \rho (\sm)$ and $\potm \equiv \pot (\sm)$, with
\sm\ the field line coordinate at the potential minimum\footnote{Here
and throughout we use the subscript ``m'' to denote quantities
evaluated on the accumulation surface $s=\sm$.}. In our application to
magnetic hot stars, we assume radiative equilibrium should set the
circumstellar temperature to be near the stellar effective
temperature. This implies a thermal energy per unit mass $kT/\mu$ that
is much smaller than the variation in the corresponding potential
energy \pot, which by eqn.~(\ref{eqn:potdef}) is typically comparable
or greater than the gravitational binding energy at the Kepler
radius. By eqn.~(\ref{eqn:density}), we thus expect that most of the
material should be confined to a relative narrow layer near the
potential minimum at $s=\sm$.

A Taylor series expansion of the potential about this minimum gives
\begin{equation} \label{eqn:taylor}
\pot(s) = \potm + \frac{(s - \sm)^{2}}{2} \potm'' + \ldots,
\end{equation}
where we have used the fact that, by definition, $\potm' = 0$.
Accordingly, in the neighborhood of the minimum, the density
distribution~(\ref{eqn:density}) may be well approximated by
\begin{equation} \label{eqn:density-approx}
\rho(s) \approx \rhom \,e^{-(s - \sm)^{2}/\height^{2}},
\end{equation}
where the RRM scale height is
\begin{equation} \label{eqn:height}
\height 
= \sqrt{\frac{2 k T/\mu}{ \potm''}}
= \sqrt{\frac{ 2 k T/\mu}{G\Mstar/\rk}} \, \sqrt{\frac{1}{\dpotm''}}.
\end{equation}
In the latter equality, the first square root is of the ratio between
the thermal energy and the gravitational binding energy at the Kepler
radius, while the second root gives the curvature length of the
effective potential. Overall, we thus see that the effect of a finite
gas pressure is to support material in a nearly Gaussian
stratification on either side of an accumulation surface.

A similar Gaussian stratification is found in models for Keplerian
disks; however, the scale height in that case grows as the
three-halves power of the distance from disk centre
\citep[\eg,][]{HumHan1997}, leading to a disk that flares outward with
increasing radius. In contrast, the RRM scale height approaches a
constant value far from the origin; for example, in the case of the
aligned dipole configuration, eqn.~(\ref{eqn:dpot-2-deriv}) gives
$\dpotm'' = 3/\rk^{2}$ when $\dr^{3} \gg 1$, yielding the asymptotic
scale height
\begin{equation} \label{eqn:height-align}
\height \rightarrow 
\rk \sqrt{\frac{2 kT/\mu} {3 G\Mstar/\rk}} \qquad (r \gg \rk) \,
\end{equation}
\citep[compare with eqn.~5 of ][]{Nak1985}. Since the ratio of thermal
energy to gravitational binding energy is typically very small, we
thus see that the disk is indeed geometrically thin, with a scale
height \height\ that is much smaller than the characteristic Kepler
radius \rk.

A convenient way to characterize the properties of such a thin
accumulation layer is in terms of its \emph{surface} density. For
field line with a projection cosine \mubm\ to the surface normal, the
associated local surface density \sigmam\ can be obtained by
integration of eqn.~(\ref{eqn:density-approx}) over the Gaussian
hydrostatic stratification,
\begin{equation}\label{eqn:sigm}
\sigmam \approx \sqrt{\pi} \rhom \mubm \height \, .
\end{equation}
For the application here to magnetic hot stars, we next develop a
specific model for the global distribution of this surface density as
being proportional to the rate of material build-up by the loading
from the star's radiatively driven wind.


\section{Mass Loading of Accumulation Surfaces}\label{sec:plasma}

\subsection{The Accumulation Rate for Surface Density} \label{sec:accum-rate}

The high luminosity of hot stars is understood to give rise to a
radiatively driven stellar wind; in a magnetic hot star, this wind
provides a key mechanism to load mass into the effective
gravito-centrifugal potential wells around the accumulation surfaces.

For a radiatively driven wind in the presence of a magnetic field,
\citet{OwoudD2004} derive an expression for the mass flux density at
the stellar surface, in terms of the spherical mass loss rate \Mdot\
predicted by the standard CAK wind model \citep{Cas1975}. For a dipole
flux-tube bundle intersecting the stellar surface $r=\Rstar$ with a
projection cosine \mubs, and having a cross-sectional area \dAs, the
rate of mass increase is
\begin{equation}
\mdot\ = 
\frac{2 \mubs \Mdot}{4\pi\Rstar^{2}} \, \dAs \, ,
\end{equation}
where the factor 2 accounts for the mass injection from two distinct
footpoints.

In a highly supersonic stellar wind, the collision of material from
opposite footpoints leads to strong shocks that heat the plasma
initially to temperatures of millions of degrees; this dogma, advanced
by \citet{BabMon1997a,BabMon1997b} in their MCWS paradigm, has been
amply confirmed both by MHD modeling \citep[\eg,][]{udDOwo2002}, and
by analysis of the observed X-ray emission from the superheated
post-shock plasma \citep[\eg,][]{Gag2004}. Eventually, the plasma
cools radiatively back to temperatures near the stellar effective
temperature of a few times $10^{4}\,\kelv$. As discussed already in
Sec.~\ref{sec:hydstrat}, at such temperatures material trapped within
the effective potential well will quickly settle into a relatively
narrow hydrostatic stratification centred on the potential minima.

For the moment, let us consider the relatively simple, common case
that there is a single minimum at field line coordinate \sm, at which
point the flux tube has area \dAm\ and intersection cosine \mubm\ with
the accumulation surface normal. Then the associated rate of increase
in surface density \sigmam\ can be written
\begin{equation}\label{eqn:sigdota}
\sigmamdot = \mdot \mubm/\dAm =
\mubm \, \mubs \, \frac{\Mdot}{2 \pi \Rstar^{2}} \,\frac{\dAs}{\dAm} \, .
\end{equation}
From the conservation of magnetic flux, $\nabla \cdot \vmag = 0$, we
have
\begin{equation}
\dAs \Bstar = \dAm \Bm \, ,
\end{equation}
which when applied to eqn.~(\ref{eqn:sigdota}) gives
\begin{equation}\label{eqn:sigdotb}
\sigmamdot = 
\mubs \, \mubm \, \frac{\Mdot}{2 \pi \Rstar^{2}} \,\frac{\Bm}{\Bstar} \, .
\end{equation}
For a dipole field, this declines with radius $r$ as $\sigmamdot \propto B \propto r^{-3}$.

\subsection{The Time Evolution of the Volume Density}

The above merely gives the \emph{rate} at which surface density
increases in the accumulation surface. In the idealized limit that the
field is arbitrarily strong, the actual surface density could thus
increase without bound. In reality, for any large but finite field,
the finite magnetic tension could only contain a finite mass of
material. As analyzed in the Appendix, above some \emph{breakout}
density the net centrifugal and gravitational force should overwhelm
the tension, leading to centrifugal ejection that effectively empties
the accumulation surface.

This view suggests a simple model in which the local surface density
builds linearly with the time $t$ since the last evacuation, $\sigmam
= \sigmamdot t$. Applying such a model to eqns~(\ref{eqn:sigm})
and~(\ref{eqn:sigdotb}), we can eliminate the normalizing term \rhom\
from the equation~(\ref{eqn:density}) for the \emph{volume} density,
to find the time evolution of this density as
\begin{equation} \label{eqn:density-full}
\rho(s,t) \approx 
\frac{\Mdot t \,\mubs}{2 \pi^{3/2} \Rstar^{2} \, \height} 
\,\frac{\Bm}{\Bstar}\, e^{-\mu [\pot(s) - \potm]/k T} \, .
\end{equation}
For field lines that exhibit \emph{two} potential minima, we divide
the latter expression by the factor
\begin{equation}
f = 1 + \frac{\heightp}{\height}\,
\frac{\Bm}{\Bmp}\, e^{\mu [\potm - \potmp]/kT},
\end{equation}
where \heightp, \Bmp\ and \potmp\ are the scale height, field strength
and potential evaluated at the secondary minimum with field line
coordinate \smp. This factor accounts for the partitioning of plasma
between the two minima, assuming a free exchange of material leads to
a common hydrostatic stratification.

We emphasize that eqn.~(\ref{eqn:density-full}) applies only to field
lines that intersect one or more accumulation surfaces, with one or
more potential minima. For all other field lines we set the density to
zero, reflecting the notion that there can be no stable accumulation
of material over time. Moreover, even for field lines intersecting a
surface, we only apply eqn.~(\ref{eqn:density-full}) up to the
bracketing \emph{maxima} in the potential. Beyond these maxima, the
decreasing potential suggests an exponentially growing density,
contrary to the true physical picture of no accumulation. We resolve
this difficulty by setting the density $\rho$ to zero in the regions
beyond the potential maxima\footnote{Unless, of course, a region
beyond a potential maximum belongs to a neighbouring secondary minimum
at $s=\smp$ -- in which case, the expression for the density remains
valid.}. This approach creates discontinuities in the densities at the
maxima themselves, but in most cases ones so small that they are
unlikely to be of much significance for the overall model.


\section{Circumstellar Emission} \label{sec:emission}

\begin{figure*}
\epsffile{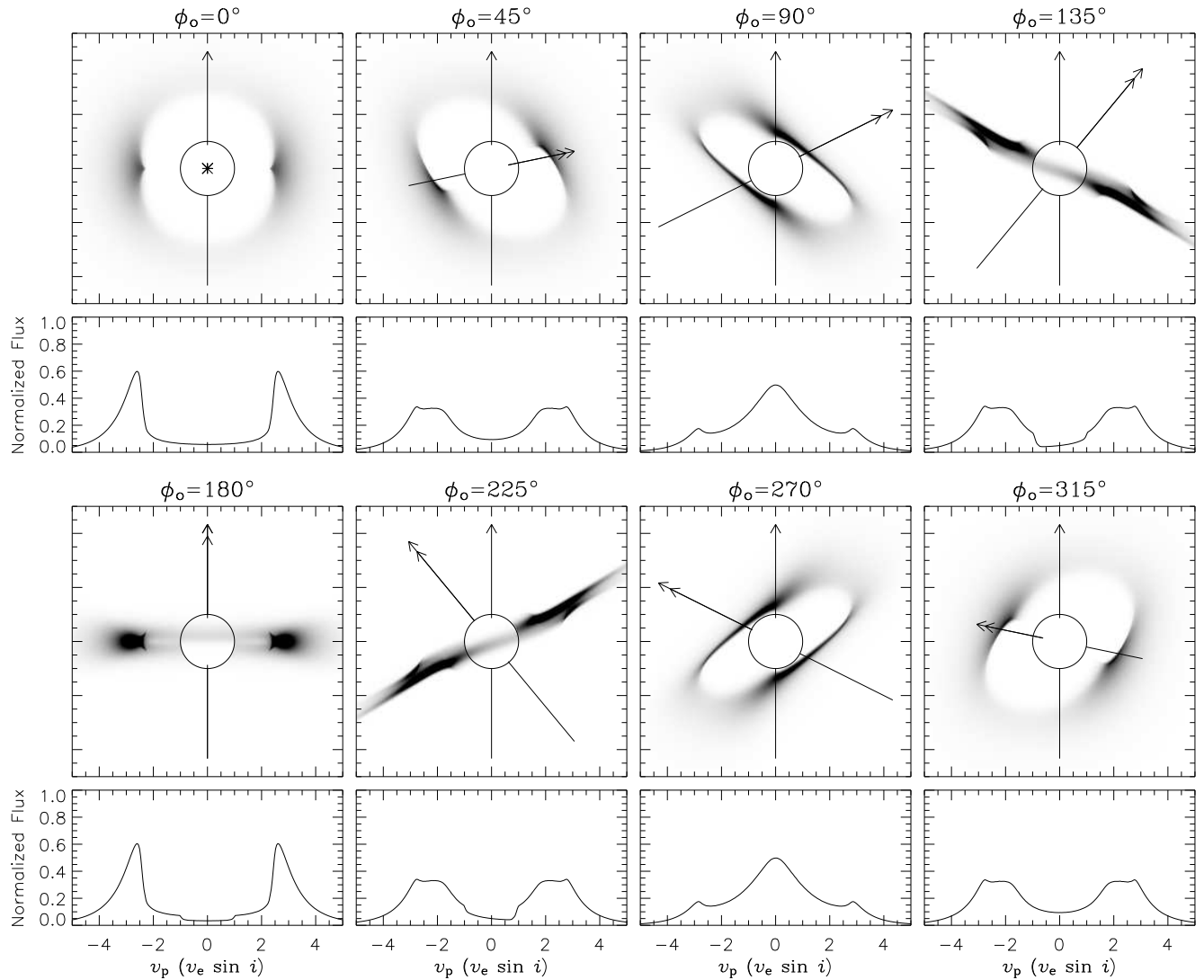}
\caption{Maps of the emission from the $\beta=60\degr$ RRM model (see
text), for eight different values of the observer azimuth \phio;
darker shading indicates greater intensity. In each map, the rotation
and magnetic axes are shown as a single- and double-headed arrows,
while the outline of the central star (whose contributions toward the
emission are neglected) is shown by a circle. Beneath each map, we
plot the corresponding emission spectrum as a function of the
projected velocity \velproj, the latter being measured in units of the
projected equatorial rotation velocity \velesini\ of the
star.}\label{fig:emission}
\end{figure*}

Let us now apply the expression~(\ref{eqn:density-full}) for the
density distribution toward calculating the circumstellar line
emission from a tilted-dipole RRM configuration. As a typical example
case, we adopt a tilt angle $\beta=60\degr$, and set the stellar
angular velocity $\Omega$ to 50 percent of the critical rate,
\begin{equation}
\srotc \equiv \sqrt{\frac{8 G \Mstar}{27 \Rstar^{3}}} \, ,
\end{equation}
at which the surface centrifugal force at the equator would balance
gravity\footnote{Note that, while this expression for \srotc\ is
appropriate to a centrifugally-distorted star, we have assumed
elsewhere, for simplicity, that the star remains spherical. For the
particular choice $\srot = 0.5\,\srotc$, the stellar oblateness
remains small, with the equatorial surface radius being barely 4
percent larger than that at the poles.}; this rotation rate
corresponds to a dimensionless stellar radius $\drstar \equiv
\Rstar/\rk = 0.42$. Our choice of parameters is loosely guided by the
rotation rate and magnetic tilt inferred from the \citet{GroHun1997}
model for the surface of \sOriE; however, we make no attempt at fine
tuning, since we are more concerned here with demonstrating the
capabilities of the RRM model, than with obtaining an accurate picture
of this particular helium-strong star.

To specify the temperature $T$ and mean molecular weight $\mu$ in the
model, we introduce the dimensionless quantity
\begin{equation}
\eratio \equiv \frac{k T \Rstar}{\mu G \Mstar},
\end{equation}
which characterizes the ratio of thermal to gravitational binding
energy at the stellar surface. In the photospheres of early-type
stars, this ratio is of the order $\sim 10^{-3}$. Following again the
scenario that the circumstellar environment remains at a temperature
close to photospheric (\cf\ Sec.~\ref{sec:hydstrat}), we therefore set
$\eratio = 10^{-3}$ throughout.

We now make the assumption that the plasma volume emissivity \jw, at a
wavelength \wave, may be characterized by the relation
\begin{equation}
\jw = \jz \, \rho^{2} \, \delta(\wave - \wavez);
\end{equation}
here, \jz\ and \wavez\ are constants, and $\delta(\ldots)$ is the
Dirac delta function. This expression is intended to mirror the
process of monochromatic line emission at a rest-frame wavelength
\wavez, arising from the density-squared radiative recombination of
ionized hydrogen. Integrating the emissivity along a given ray passing
through the magnetosphere, the observed surface intensity of the
emission is given by
\begin{equation}
\isurfw = \jz \, \int_{0}^{\infty} \rho^{2}(\zobs) \, \delta(\wave -\wavez[1 +
\velproj(\zobs)/c]) \, \diff\zobs,
\end{equation}
where \zobs\ is the distance along the ray from the observer,
\velproj\ is the projection of the local plasma velocity onto the ray,
and $c$ is the speed of light. For rays intersecting the star, this
integral must be truncated at the stellar surface, to account for the
occultation of radiation incident from the star's far side. Note that
this simple expression does not include the emission of radiation by
the star itself; therefore, it should be taken to represent the
notional circumstellar component of the net radiation from the system,
with the understanding that the corresponding photospheric component
has been subtracted away. Of course, such an interpretation is in
itself an approximation, since our emission model does not account for
episodes when circumstellar plasma transits the disk and absorbs
stellar radiation over the range $-\velesini < \velproj < \velesini$;
however, at the level of the present investigation, this approximation
is entirely sufficient.

Because the magnetospheric plasma co-rotates rigidly with the star,
\velproj\ may be expressed as
\begin{equation}
\velproj(\zobs) = \imp \, \velesini,
\end{equation}
where the impact parameter \imp\ is the perpendicular distance, in
units of \Rstar, between the ray and the rotation axis, and \vele\ is
the equatorial rotation velocity of the star. These quantities, and
therefore \velproj\ too, are independent of \zobs; hence, the
intensity may be written in the form
\begin{equation} \label{eqn:intensity}
\isurfw = \jz \, 
\delta(\wave - \wavez[1 + \imp\,\velesini/c]) \, \int_{0}^{\infty} 
\rho^{2}(\zobs) \, \diff \zobs.
\end{equation}
From this latter expression, it can be seen that all plasma having the
same \imp\ -- that is, situated on a plane parallel to the rotation
axis -- will radiate monochromatically at the same wavelength.

Applying eqn.~(\ref{eqn:intensity}) to the known density distribution
of the $\beta=60\degr$ RRM configuration, Fig.~\ref{fig:emission}
shows maps of the wavelength-integrated emission intensity
\begin{equation}
\isurf \equiv \int_{0}^{\infty} \isurfw \, \diff\lambda,
\end{equation}
extending out to 5\,\Rstar\ in directions parallel and perpendicular
to the projected rotation axis. The observer is situated at the same
inclination $\incl=60\degr$ that we adopt in Sec.~\ref{sec:surfaces},
and at eight differing values of the azimuth \phio, separated by
uniform increments of $45\degr$. Beneath each map we show the
corresponding spectrum, in which the spatially-integrated emission is
plotted as a function of the projected velocity \velproj. Since we are
interested more in the distribution of emission than its absolute
value, Fig.~\ref{fig:emission} adopts an arbitrary (although
consistent) normalization for the intensities in both the maps and the
spectra.

In Section~\ref{sec:surfaces}, we demonstrate that the accumulation
surface for a $\beta=60\degr$ tilted dipole takes the form of a warped
disk. The emission maps in Fig.~\ref{fig:emission} reveal that the
distribution of material across this disk is decidedly
non-uniform. Specifically, the distribution is dominated by two
clouds, located near the inner edge of the disk at the intersection
between the rotational and magnetic equators. Seen from an inertial
frame, these clouds appear to rotate synchronously with the star;
furthermore, their characteristic twin-peaked emission spectrum
displays temporal variations in the form of a double S-wave.

These findings exhibit an encouraging degree of agreement with the
inferred behaviour of \sOriE. As discussed in the introduction,
observations of this star indicate that circumstellar plasma is
concentrated at the intersection between rotational and magnetic
equators \citep{GroHun1982,Bol1987,ShoBol1994}. Without the need for
any special tuning, beyond the requirement that the tilt angle $\beta$
be moderate, the RRM model naturally reproduces such a
distribution. It also accounts for the Balmer line emission from the
star, which -- when the measured longitudinal field is strongest,
corresponding to $\phio=0\degr$ and $\phio=180\degr$ in
Fig.~\ref{fig:emission} -- is observed to exhibit strong peaks
situated at $\velproj \approx \pm 3 \velesini$ (Groote, personal
communication).


\section{Discussion}\label{sec:discussion}

As we discuss in Sec.~\ref{sec:intro}, there have been a number of
previous studies that have made use of the rigid-field approach
(\cf~Sec.~\ref{sec:model}) to determining the regions where
circumstellar material can accumulate (Sec.~\ref{sec:surfaces}). The
present paper builds on these studies, by presenting a
physically-grounded RRM model for the steady accumulation of wind
plasma in the circumstellar environment (Secs.~\ref{sec:hydstrat}
\&~\ref{sec:plasma}), that is able to make specific predictions
regarding the observables associated with this plasma
(Sec.~\ref{sec:emission}).

Our treatment of the mass loading of accumulation surfaces differs
markedly from the approach advanced by \citet{Nak1985}, who fixed the
plasma density at each point by requiring equal magnetic and kinetic
energy densities (see his eqn.~13); such a choice is guided by the
notion that when the density is high enough for the kinetic energy due
to rotation to dominate the magnetic energy, the field lines break
open, and any subsequently-added plasma leaks away from the system. By
contrast, our approach to deriving the plasma distribution focuses on
the mass accumulation rather than on leakage
(\cf~Sec.~\ref{sec:plasma}). While \citet{Nak1985} treats the mass
leakage as a gradual, quasi-steady process, we view it more as an
episodic evacuation caused by magnetic breakout (see Appendix), which
effectively resets the mass accumulation.

In actual systems the mass distribution may reflect a combination of
both perspectives, and there are even other alternative frameworks for
treating the problem \citep[see,
\eg,][]{MicStu1974,HavGoe1984}. However, one particularly favourable
aspect of the present model is that it can naturally reproduce the
plasma concentrations at the intersection of rotational and magnetic
equators, as is inferred from observations of \sOriE. To obtain a
similar result, \citet{Nak1985} had to make the \emph{ad hoc}
assumption that some process of diffusion, across magnetic field
lines, leads to the redistribution of plasma into the desired
configuration.

We turn now to a brief discussion of the recent paper by
\citet{Pre2004}, which was published during the final stages of
preparation of the present work. These authors found the same
accumulation surfaces as we present in Sec.~\ref{sec:surfaces}, but
using the alternative formulation based around consideration of the
loci where all forces tangential to field lines vanish. They were the
first to discover the possibility of leaves and the truncated cones
occurring at $\beta=90\degr$, and even explored the case of tilted
dipoles offset from the origin. However, \citet{Pre2004} stopped short
of considering the mass loading of the accumulation surfaces, and
instead focused on their geometrical form. As such, their study does
not make the kind of specific predictions for observable emission line
variations that we provide in Sec.~\ref{sec:emission}.

Another relevant comparison here is to the \emph{Magnetically Torqued
Disk} (MTD) model proposed by \citet{Cas2002} to explain the
circumstellar emission of Be stars. Building upon insights from
one-dimensional equatorial plane models of magnetically torqued
stellar winds \citep{WebDav1967,BelMac1976}, this analysis centres on
an assumed empirical scaling of the azimuthal velocity, which
initially increases as a rigid-body law out to some peak, and then
declines asymptotically with angular momentum conservation. When this
peak occurs above the Kepler radius, the model envisions that the
associated torquing of the wind outflow can lead to formation of a
`quasi-Keplerian' disk. But recent dynamical simulations
\citep{OwoudD2003} indicate that fields marginally strong enough to
spin wind material beyond Keplerian rotation tend instead to lead to
centrifugal mass ejection rather than a Keplerian disk.

For much stronger fields, the region of rigid rotation becomes more
extended, and the MTD scenario can be viewed as becoming similar to
the field-aligned rotation case ($\beta=0$) of the RRM model developed
here\footnote{Note, however, that while the MTD analysis emphasized
the torquing role of the magnetic field, in the RRM model the rigid
field also plays a crucial role in holding material down against a net
centrifugal force that, for radii beyond the Kepler radius, exceeds
the inward force of gravity.}. Although the disk rotation is
rigid-body rather than Keplerian, the tendency for the bulk of the
material to build up in the region near the Kepler radius means that
the resulting line emission should develop a doubled-peaked profile
that might be quite difficult to distinguish from what is expected
from a Keplerian disk. Note, however, that such rigid-body disks seem
unlikely to produce the long-term (years to decade) violet/red (V/R)
variations often observed in Be-star emission lines \citep{Tel1994};
such variations seem instead likely to be the result of long-term
precession of elliptical orbits within a Keplerian disk
\citep{SavHee1993}. As such, we do not believe that the RRM model is
likely to be of general relevance to explaining Be star emission.
However, as noted above, it does seem quite well-suited to explaining
the rotationally modulated emission of Bp stars like \sOriE.

On a concluding note, we draw attention to the fact that X-ray flaring
has been detected in \sOriE\ by \emph{ROSAT} \citep{GroSch2004}, and
subsequently by \emph{XMM-Newton} \citep{San2004}. \citet{Mul2004} has
argued that the flares originate from the B2 star itself, rather than
from an unseen low-mass companion. If this is indeed the case, then we
suggest a likely mechanism for the flare generation is thermal heating
arising from magnetic reconnection. As we discuss in the Appendix, we
expect the outer parts of the accumulation surface to undergo
relatively frequent breakout events, during which stressed magnetic
field lines will reconnect and release significant quantities of
energy. We intend to explore this hypothesis further in a future paper
\citep[and see also][]{udD2004}.


\section{Summary} \label{sec:summary}

We have presented a new Rigidly Rotating Magnetosphere model for the
circumstellar plasma distributed around magnetic early-type stars. By
assuming that field lines remain completely rigid, and co-rotate with
the star, we are able to find regions in the circumstellar environment
where plasma can accumulate under hydrostatic equilibrium; in the
general case of a tilted dipole field, these regions take the form of
a geometrically-thin, warped disk, whose mean surface normal lies
between the misaligned magnetic and rotation axes. When coupled with a
quantitative description of the accumulation process, our treatment
allows us to evaluate the density throughout the circumstellar
environment, and thereby calculate observables such as emission-line
spectra.

This RRM model shows promise; even without a fine tuning of
parameters, it reproduces the principal features of \sOriE, the
archetype of the variable-emission helium-strong stars. In a
forthcoming paper, we will investigate the extent to which the model
can reproduce the more detailed aspects of this star -- in particular,
the \emph{strength} of the emission lines, as well as their shape, and
the eclipse-line variations seen in photometric indices. We also plan
to examine whether the model can be applied to other magnetic
early-type stars.


\section*{Acknowledgments}

RHDT acknowledges partial support from the UK Particle Physics and
Astronomy Research Council. Both RHDT and SPO acknowledge partial
support from US NSF grant AST-0097983. Initial ideas for this work
began during SPO's 2003 sabbatical leave visit to the UK under PPARC
support; he particularly thanks J. Brown of University Glasgow and
A. Willis of University College London for their hospitality. We also
thank Detlef Groote for many useful discussions regarding the
observations of $\sigma$ Ori E.


\bibliography{magdisk}

\begin{thebibliography}{}

\bibitem[\protect\citeauthoryear{{Babcock}}{{Babcock}}{1958}]{Bab1958}
{Babcock} H.~W.,  1958, \apj, 128, 228

\bibitem[\protect\citeauthoryear{{Babel} \& {Montmerle}}{{Babel} \&
  {Montmerle}}{1997a}]{BabMon1997a}
{Babel} J.,  {Montmerle} T.,  1997a, \apjl, 485, 29

\bibitem[\protect\citeauthoryear{{Babel} \& {Montmerle}}{{Babel} \&
  {Montmerle}}{1997b}]{BabMon1997b}
{Babel} J.,  {Montmerle} T.,  1997b, \aap, 323, 121

\bibitem[\protect\citeauthoryear{{Baliunas}, {Donahue}, {Soon} \& {et
  al.}}{{Baliunas} et~al.}{1995}]{Bal1995}
{Baliunas} S.~L.,  {Donahue} R.~A.,  {Soon} W.~H.,    {et al.} 1995, \apj, 438,
  269

\bibitem[\protect\citeauthoryear{{Belcher} \& {MacGregor}}{{Belcher} \&
  {MacGregor}}{1976}]{BelMac1976}
{Belcher} J.~W.,  {MacGregor} K.~B.,  1976, \apj, 210, 498

\bibitem[\protect\citeauthoryear{{Bolton}, {Fullerton}, {Bohlender},
  {Landstreet} \& {Gies}}{{Bolton} et~al.}{1987}]{Bol1987}
{Bolton} C.~T.,  {Fullerton} A.~W.,  {Bohlender} D.,  {Landstreet} J.~D.,
  {Gies} D.~R.,  1987, in {Slettebak} A.,  {Snow} T.~P.,  eds, IAU Colloq. 92:
  Physics of Be Stars, Cambridge University Press, Cambridge, p.~82

\bibitem[\protect\citeauthoryear{{Cassinelli}, {Brown}, {Maheswaran}, {Miller}
  \& {Telfer}}{{Cassinelli} et~al.}{2002}]{Cas2002}
{Cassinelli} J.~P.,  {Brown} J.~C.,  {Maheswaran} M.,  {Miller} N.~A.,
  {Telfer} D.~C.,  2002, \apj, 578, 951

\bibitem[\protect\citeauthoryear{{Castor}, {Abbott} \& {Klein}}{{Castor}
  et~al.}{1975}]{Cas1975}
{Castor} J.~I.,  {Abbott} D.~C.,    {Klein} R.~I.,  1975, \apj, 195, 157

\bibitem[\protect\citeauthoryear{{Del Zanna} \& {Mason}}{{Del Zanna} \&
  {Mason}}{2003}]{ZanMas2003}
{Del Zanna} G.,  {Mason} H.~E.,  2003, \aap, 406, 1089

\bibitem[\protect\citeauthoryear{{Donati}, {Babel}, {Harries}, {Howarth},
  {Petit} \& {Semel}}{{Donati} et~al.}{2002}]{Don2002}
{Donati} J.-F.,  {Babel} J.,  {Harries} T.~J.,  {Howarth} I.~D.,  {Petit} P.,
   {Semel} M.,  2002, \mnras, 333, 55

\bibitem[\protect\citeauthoryear{{Gagn\'{e}}, {Oksala}, {Cohen}, Tonnesen,
  {ud-Doula}, {Owocki} \& {MacFarlane}}{{Gagn\'{e}} et~al.}{2004}]{Gag2004}
{Gagn\'{e}} M.,  {Oksala} M.,  {Cohen} D.~H.,  Tonnesen S.~K.,  {ud-Doula} A.,
  {Owocki} S.~P.,    {MacFarlane} J.~J.,  2004, \apj, in preparation

\bibitem[\protect\citeauthoryear{{Groote} \& {Hunger}}{{Groote} \&
  {Hunger}}{1982}]{GroHun1982}
{Groote} D.,  {Hunger} K.,  1982, \aap, 116, 64

\bibitem[\protect\citeauthoryear{{Groote} \& {Hunger}}{{Groote} \&
  {Hunger}}{1997}]{GroHun1997}
{Groote} D.,  {Hunger} K.,  1997, \aap, 319, 250

\bibitem[\protect\citeauthoryear{{Groote} \& {Schmitt}}{{Groote} \&
  {Schmitt}}{2004}]{GroSch2004}
{Groote} D.,  {Schmitt} J.~H.~M.~M.,  2004, \aap, 418, 235

\bibitem[\protect\citeauthoryear{{Havnes} \& {Goertz}}{{Havnes} \&
  {Goertz}}{1984}]{HavGoe1984}
{Havnes} O.,  {Goertz} C.~K.,  1984, \aap, 138, 421

\bibitem[\protect\citeauthoryear{{Henrichs}, {de Jong}, {Donati} \& {et
  al.}}{{Henrichs} et~al.}{2000}]{Hen2000}
{Henrichs} H.~F.,  {de Jong} J.~A.,  {Donati} J.-F.,    {et al.} 2000, in
  {Smith} M.~A.,  {Henrichs} H.~F.,   {Fabregat} J.,  eds, ASP Conf. Ser. 214:
  IAU Colloq. 175: The Be Phenomenon in Early-Type Stars, p.~324

\bibitem[\protect\citeauthoryear{{Hesser}, {Walborn} \& {Ugarte}}{{Hesser}
  et~al.}{1976}]{Hes1976}
{Hesser} J.~E.,  {Walborn} N.~R.,    {Ugarte} P.~P.,  1976, \nat, 262, 116

\bibitem[\protect\citeauthoryear{{Hummel} \& {Hanuschik}}{{Hummel} \&
  {Hanuschik}}{1997}]{HumHan1997}
{Hummel} W.,  {Hanuschik} R.~W.,  1997, \aap, 320, 852

\bibitem[\protect\citeauthoryear{{Landstreet} \& {Borra}}{{Landstreet} \&
  {Borra}}{1978}]{LanBor1978}
{Landstreet} J.~D.,  {Borra} E.~F.,  1978, \apjl, 224, 5

\bibitem[\protect\citeauthoryear{{Michaud}, {Charland} \&
  {Megessier}}{{Michaud} et~al.}{1981}]{Mic1981}
{Michaud} G.,  {Charland} Y.,    {Megessier} C.,  1981, \aap, 103, 244

\bibitem[\protect\citeauthoryear{{Michel} \& {Sturrock}}{{Michel} \&
  {Sturrock}}{1974}]{MicStu1974}
{Michel} F.~C.,  {Sturrock} P.~A.,  1974, \planss, 22, 1501

\bibitem[\protect\citeauthoryear{{Mullan}}{{Mullan}}{2004}]{Mul2004}
{Mullan} D.~J.,  2004, \aap, in preparation

\bibitem[\protect\citeauthoryear{{Nakajima}}{{Nakajima}}{1985}]{Nak1985}
{Nakajima} R.,  1985, \apss, 116, 285

\bibitem[\protect\citeauthoryear{{Neiner}, {Geers}, {Henrichs}, {Floquet},
  {Fr{\' e}mat}, {Hubert}, {Preuss} \& {Wiersema}}{{Neiner}
  et~al.}{2003}]{Nei2003}
{Neiner} C.,  {Geers} V.~C.,  {Henrichs} H.~F.,  {Floquet} M.,  {Fr{\' e}mat}
  Y.,  {Hubert} A.-M.,  {Preuss} O.,    {Wiersema} K.,  2003, \aap, 406, 1019

\bibitem[\protect\citeauthoryear{{Owocki} \& {ud-Doula}}{{Owocki} \&
  {ud-Doula}}{2003}]{OwoudD2003}
{Owocki} S.~P.,  {ud-Doula} A.,  2003, in {Balona} L.~A.,  {Henrichs} H.~F.,
  {Medupe} R.,  eds, ASP Conf. Ser. 305: Magnetic Fields in O, B and A Stars,
  p.~350

\bibitem[\protect\citeauthoryear{{Owocki} \& {ud-Doula}}{{Owocki} \&
  {ud-Doula}}{2004}]{OwoudD2004}
{Owocki} S.~P.,  {ud-Doula} A.,  2004, \apj, 600, 1004

\bibitem[\protect\citeauthoryear{{Pedersen}}{{Pedersen}}{1979}]{Ped1979}
{Pedersen} H.,  1979, \aaps, 35, 313

\bibitem[\protect\citeauthoryear{{Pedersen} \& {Thomsen}}{{Pedersen} \&
  {Thomsen}}{1977}]{PedTho1977}
{Pedersen} H.,  {Thomsen} B.,  1977, \aaps, 30, 11

\bibitem[\protect\citeauthoryear{{Preuss}, {Sch{\" u}ssler}, {Holzwarth} \&
  {Solanki}}{{Preuss} et~al.}{2004}]{Pre2004}
{Preuss} O.,  {Sch{\" u}ssler} M.,  {Holzwarth} V.,    {Solanki} S.~K.,  2004,
  \aap, 417, 987

\bibitem[\protect\citeauthoryear{{Sanz-Forcada}, {Franciosini} \&
  {Pallavicini}}{{Sanz-Forcada} et~al.}{2004}]{San2004}
{Sanz-Forcada} J.,  {Franciosini} E.,    {Pallavicini} R.,  2004, \aap, in
  press

\bibitem[\protect\citeauthoryear{{Savonije} \& {Heemskerk}}{{Savonije} \&
  {Heemskerk}}{1993}]{SavHee1993}
{Savonije} G.~J.,  {Heemskerk} M.~H.~M.,  1993, \aap, 276, 409

\bibitem[\protect\citeauthoryear{{Shore}}{{Shore}}{1993}]{Sho1993}
{Shore} S.~N.,  1993, in ASP Conf. Ser. 44: IAU Colloq. 138: Peculiar versus
  Normal Phenomena in A-type and Related Stars, p.~528

\bibitem[\protect\citeauthoryear{{Shore} \& {Brown}}{{Shore} \&
  {Brown}}{1990}]{ShoBro1990}
{Shore} S.~N.,  {Brown} D.~N.,  1990, \apj, 365, 665

\bibitem[\protect\citeauthoryear{{Shore}, {Brown}, {Sonneborn}, {Landstreet} \&
  {Bohlender}}{{Shore} et~al.}{1990}]{Sho1990}
{Shore} S.~N.,  {Brown} D.~N.,  {Sonneborn} G.,  {Landstreet} J.~D.,
  {Bohlender} D.~A.,  1990, \apj, 348, 242

\bibitem[\protect\citeauthoryear{{Short} \& {Bolton}}{{Short} \&
  {Bolton}}{1994}]{ShoBol1994}
{Short} C.~I.,  {Bolton} C.~T.,  1994, in {Balona} L.~A.,  {Henrichs} H.~F.,
  {Le Contel} J.~M.,  eds, IAU Symp. 162: Pulsation, Rotation, and Mass Loss in
  Early-Type Stars, Kluwer, Dordrecht, p.~171

\bibitem[\protect\citeauthoryear{{Telting}, {Heemskerk}, {Henrichs} \&
  {Savonije}}{{Telting} et~al.}{1994}]{Tel1994}
{Telting} J.~H.,  {Heemskerk} M.~H.~M.,  {Henrichs} H.~F.,    {Savonije} G.~J.,
   1994, \aap, 288, 558

\bibitem[\protect\citeauthoryear{{ud-Doula}}{{ud-Doula}}{2003}]{udD2003}
{ud-Doula} A.,  2003, PhD thesis, University of Delaware

\bibitem[\protect\citeauthoryear{{ud-Doula} \& {Owocki}}{{ud-Doula} \&
  {Owocki}}{2002}]{udDOwo2002}
{ud-Doula} A.,  {Owocki} S.~P.,  2002, \apj, 576, 413

\bibitem[\protect\citeauthoryear{{ud-Doula}, {Townsend} \& {Owocki}}{{ud-Doula}
  et~al.}{2004}]{udD2004}
{ud-Doula} A.,  {Townsend} R.~H.~D.,    {Owocki} S.~P.,  2004, in {Ignace} R.,
  {Gayley} K.~G.,  eds, ASP Conf. Ser., in press: The Nature and Evolution of
  Disks around Hot Stars

\bibitem[\protect\citeauthoryear{{Weber} \& {Davis}}{{Weber} \&
  {Davis}}{1967}]{WebDav1967}
{Weber} E.~J.,  {Davis} L.~J.,  1967, \apj, 148, 217

\bibitem[\protect\citeauthoryear{{Wilson}}{{Wilson}}{1978}]{Wil1978}
{Wilson} O.~C.,  1978, \apj, 226, 379

\bibitem[\protect\citeauthoryear{{Wolff} \& {Morrison}}{{Wolff} \&
  {Morrison}}{1974}]{WolMor1974}
{Wolff} S.~C.,  {Morrison} N.~D.,  1974, \pasp, 86, 935

\end{thebibliography}

\bibliographystyle{mn2e}


\appendix

\section{Breakout of Accumulated Material}\label{sec:breakout}

\subsection{Breakout Time}

In deriving the form of the accumulation surfaces, we have assumed an
arbitrarily strong, rigid field. But in practice the finite magnitude
of any stellar magnetic field means that there is a limit to the mass
that can be contained against the centrifugal force; when the density
gets too high, the material should \emph{break out} from the field
containment. Just prior to such an episode, the over-stressed magnetic
field becomes distorted from its equilibrium configuration, sagging
radially outward as it passes through the dense material in
accumulation surfaces. Under these circumstances, the inward force
arising from the tension in the distorted field lines barely balances
the net outward gravito-centrifugal force; therefore, we can construct
an approximate condition for the occurrence of breakout by equating
these two forces. Focusing our analysis in this appendix on the simple
case of an aligned dipole field ($\beta=0\degr$), the breakout
condition may be expressed as
\begin{equation}
\rhob \left[ \Omega^{2} r - \frac{G\Mstar}{r^{2}} \right] \approx
\frac{B^{2}}{4\pi \height},
\end{equation}
where \rhob\ represents a breakout value for the peak density at
radius $r$ within the equatorial plane, and the scale height \height\
appears as the typical curvature radius of the distorted magnetic
field lines, whose tension $B^{2}/4\pi$ generates the balancing inward
force. In analogy with eqn.~(\ref{eqn:sigm}) we can define an
associated breakout surface density $\sigmab \equiv \sqrt{\pi} \rhob
\height$, where we have taken $\mubm=1$ as appropriate to the
equatorial plane. Using the Kepler radius
\rk~(eqn.~\ref{eqn:r-kepler}) to scale both the local radius ($\xi
\equiv r/\rk$) and the stellar radius ($\drstar \equiv \Rstar/\rk$),
we have for the usual dipole field scaling $B \sim r^{-3}$,
\begin{equation}
\sigmab \gstar \drstar^{2} \left[ \xi - \frac{1}{\xi^{2}} \right] =
\frac{\Bstar^{2} \drstar^{6}}{4\pi \xi^{6}} \sqrt{\pi} \, ,
\end{equation}
or, solving for the breakout surface density,
\begin{equation} \label{eqn:sigma-b}
\sigmab(\xi) = \frac{\Bstar^{2} \drstar^{4}}{4\pi \gstar} \,
\frac{\sqrt{\pi}}{\xi^{4} (\xi^{3}-1)}.
\end{equation}
Scaled in terms of typical parameters for a rotating, magnetic B-star,
we find a characteristic surface density for breakout
\begin{equation}
\sigmastar \equiv \frac{\Bstar^{2} \drstar^{4}}{4\pi \gstar} 
\approx 8 \,\gcmcm \, \frac{B_{3}^{2} \drstar^{4}}{g_{4}} \, ,
\end{equation}
where $B_{3} \equiv B/10^{3}\,\gauss$ and $g_{4} \equiv
\gstar/10^{4}\,\cmss$.

For comparison, note that for a dipole field the surface density
accumulation rate in eqn.~(\ref{eqn:sigdotb}) has the scaling
\begin{equation}\label{eqn:sigdotc}
\sigmamdot(\xi) = 
\mubs \, \frac{\drstar^{3}}{\xi^{3}} \, \frac{\Mdot}{2 \pi \Rstar^{2}} \,,
\end{equation}
where once more we have assumed $\mubm=1$. Then for each scaled radius
$\xi$ we can define a characteristic \emph{breakout time}, $\tb(\xi)
\equiv \sigmab/\sigmamdot$.  Casting the stellar gravity in terms of
the surface escape speed and free-fall time, $\gstar = \velesc/2\tff$,
the ratio of breakout to free-fall time becomes
\begin{equation} \label{eqn:time-b}
\frac{\tb(\xi)}{\tff} = \estar \, \frac{\sqrt{\pi}}{\mubs} \, \frac{\drstar}{\xi(\xi^{3}-1)} \, .
\end{equation}
Here we have collected dimensional quantities in terms of a single,
dimensionless \emph{magnetic confinement parameter} for the
accumulation surface\footnote{Note that this is closely related to the
`\emph{wind} magnetic confinement parameter' defined by
\citet{udDOwo2002}, differing only by the order-unity substitutions
$\velinf \rightarrow \velesc$ and $\Beq \rightarrow \Bstar$, where
\velinf\ is the wind terminal speed, and \Beq\ is the stellar surface
field at the magnetic equator.},
\begin{equation} \label{eqn:eta-star}
\estar \equiv \frac{\Bstar^{2}\Rstar^{2}}{\Mdot \velesc} \approx 1.6
\times 10^{6} \, \frac{B_{3}^{2} R_{12}^{2}}{\Mdten \, v_{8}} \, ,
\end{equation}
with the latter equality giving a characteristic value in terms of
scaled parameters $R_{12} \equiv \Rstar/10^{12}\,\cm$, $\Mdten \equiv
\Mdot/10^{-10}\,\msunyr$, and $v_{8} \equiv \velesc/10^{8}\,\cms$.
Noting that the free fall time $\tff = \velesc/2\gstar = 10^{4}\,{\rm s}
\, (v_{8}/2g_{4})$, the breakout time evaluates to
\begin{equation} \label{eqn:time-b-val}
\tb(\xi) \approx 250\,\yr \,
\frac{B_{3}^{2} R_{12}^{2} \drstar}{\Mdten \, g_{4}} \, \frac{1}{\xi(\xi^{3}-1)} \, ,
\end{equation}
where we have taken $\sqrt{\pi}/\mubs \approx 2$. As a typical
example, corresponding roughly to values appropriate to \sOriE, let us
take $\Mdten = g_{4} = 1$, $B_{3} = 10$, and $R_{12} = \drstar = 1/2$,
yielding a value 12.5 for the ratio factor in
eqn.~(\ref{eqn:time-b-val}). At a location equal to twice the Kepler
radius, we then find a typical breakout time of $\tb(2) \approx
220\,\yr$.

\subsection{Mass in Accumulation Surface}

Let us next estimate the total mass in the equatorial accumulation
surface after some elapsed time $t$ since it was last emptied. The
mass contained between inner radius $r_{\rm i}= \xi_{\rm i} \rk$ and
outer radius $r_{\rm o}=\xi_{\rm o} \rk$ is given by the integral
\begin{align} \nonumber
m(t) &\approx 2 \pi \rk^{2} \, t \int_{\xi_{\rm i}}^{\xi_{\rm o}} 
\sigmamdot(\xi) \, \xi \, \diff\xi \\ \label{eqn:mass-t-int}
&\approx
\Mdot\, t \, \mubs\, \drstar \, \left( \frac{1}{\xi_{\rm i}} - \frac{1}{\xi_{\rm o}} \right) \, ,
\end{align}
where the latter equality uses eqn.~(\ref{eqn:sigdotc}).
Approximating the inner radius by the Kepler radius, $\xi_{\rm i}
\approx 1$, we then find
\begin{equation} \label{eqn:mass-t}
m(t) \approx \Mdot\, t \, \mubs\, \drstar \, \frac{\xi_{\rm o} - 1}{\xi_{\rm o}}
\approx \frac{\Mdot\, \tff \, \estar \drstar^{2} \sqrt{\pi}}{\xi_{\rm o}^{2}(\xi_{\rm o}^{2} + \xi_{\rm o} +1)} \, .
\end{equation}
The latter equality assumes the outer radius is limited by breakout,
and uses eqn.~(\ref{eqn:time-b}) to eliminate the explicit appearance
of $t$ in terms of the time-variable outer radius $\xi_{\rm o} (t)$.
Over a long time the outer radius approaches the inner (Kepler) radius
$\xi_{\rm o} \rightarrow 1$, with the total asymptotic disk mass
approaching
\begin{equation} \label{eqn:mass-inf}
m_{\infty} \approx \frac{\Mdot\, \tff \, \estar \, \drstar^{2} \sqrt{\pi}}{3}
\approx \frac {\Bstar^{2} \Rstar^{2} \drstar^{2} \sqrt{\pi}}{6 \gstar} \, ,
\end{equation}
where the latter equality uses the definition~(\ref{eqn:eta-star}) to
eliminate the confinement parameter, mass loss rate, and terminal
speed, and we have also eliminated the escape speed \velesc\ in favor
of the surface gravity \gstar. In terms of scaled parameters, this
evaluates to
\begin{equation} \label{eqn:mass-inf-val}
m_{\infty} \approx 1.5 \times 10^{-8} \msun \, \frac {B_{3}^{2} R_{12}^{2} \drstar^{2}}{g_{4}} \,.
\end{equation}
Again adopting the above typical parameters for \sOriE\ -- $B_{3}=10$,
$\drstar=R_{12}=1/2$ and $g_{4}=1$ -- we find $m_{\infty} \approx 9.3
\times 10^{-8}\,\msun$.

If instead we consider a time when the outer radius happens to be at
twice the Kepler radius, $\xi_{\rm o} = 2$, then by
eqn.~(\ref{eqn:mass-t}) the total mass is reduced by an extra factor
of $3/28 = 0.107$. For \sOriE, this now gives a total mass $m \approx
9.9 \times 10^{-9}\,\msun$. As noted above, the associated breakout
time for this twice-Kepler outer radius is about $\tb(2) \approx
220\,\yr$.

Overall, the picture from this analysis is that the outer parts of the
accumulation surface should be subject to relatively frequent breakout
events that empty mass from those regions. Over a longer time, rarer
breakouts can occur from closer in, eventually even quite near the
Kepler radius. This simple analysis formally assigns an arbitrarily
long build-up time, and thus arbitrarily large mass build-up, to the
Kepler radius itself. However, based on MHD simulations done so far
\citep{OwoudD2003}, it seems more likely that breakouts sufficiently
close to the Kepler radius, \ie\ with $\xi \lesssim 2$, should be
associated with a broader disruption of the overall field structure.
This can lead to an emptying of mass throughout the accumulation
surface, including the region around the Kepler radius itself.
Following such a \emph{global} evacuation, the relative distribution
of material at any given time is proportional to the wind accumulation
rate, as detailed in Sec.~\ref{sec:plasma}. 


\label{lastpage}

\end{document}